\documentclass[11pt]{article}
\usepackage[a4paper,margin=0.90in]{geometry}
\usepackage[T1]{fontenc}
\usepackage[utf8]{inputenc}
\usepackage{lmodern}
\usepackage{amsmath,amssymb,bm,amsfonts}
\usepackage{authblk}
\usepackage{booktabs,tabularx,array,longtable}
\usepackage{float}
\usepackage{enumitem}
\usepackage{xcolor,cite}
\usepackage{graphicx}
\usepackage{tikz}
\usepackage{fontawesome5}
\usepackage[most]{tcolorbox}
\usetikzlibrary{arrows.meta,positioning,fit,backgrounds,calc}
\usepackage[colorlinks=true,linkcolor=blue!60!black,
  urlcolor=blue!60!black,citecolor=blue!60!black]{hyperref}

\definecolor{agentblue}{HTML}{DCEFFD}
\definecolor{toolgreen}{HTML}{DFF2E1}
\definecolor{storegold}{HTML}{FFF1C9}
\definecolor{guardred}{HTML}{F9DDDD}
\newcommand{\code}[1]{\texttt{#1}}

\definecolor{blue_munsell}{rgb}{0.0, 0.5, 0.69}
\usepackage{xspace}
\newcommand{\PD}{\texttt{proton\_agent}\xspace}

\definecolor{agentblue}{RGB}{205,225,245}
\definecolor{toolgreen}{RGB}{210,240,215}
\definecolor{guardred}{RGB}{245,215,215}

\usepackage[most]{tcolorbox}
\usepackage{xcolor}
\usetikzlibrary{positioning,calc,fit,backgrounds,arrows.meta,backgrounds}

\newenvironment{User}
  {\par\medskip
   \begin{tcolorbox}[
      enhanced,
      breakable,
      colback=magenta!8,
      colframe=magenta!70!black,
      boxrule=1pt,
      arc=2mm,
      left=8pt,right=8pt,top=2pt,bottom=2pt,
      fontupper=\small,
      before upper={%
  {\large\textcolor{magenta!70!black}{\faUserCircle}}%
  \hspace{6pt}%
  {\normalsize\bfseries User:}%
  \par\smallskip
},
   ]}
  {\end{tcolorbox}\medskip}

\newenvironment{ProtonAgent}
  {\par\medskip
   \begin{tcolorbox}[
      enhanced,
      breakable,
      colback=cyan!4,
      colframe=cyan!50!black,
      boxrule=1pt,
      arc=2mm,
      left=8pt,right=8pt,top=2pt,bottom=2pt,
      fontupper=\small,
      before upper={%
        {\large\textcolor{cyan!50!black}{\faRobot}}%
        \hspace{6pt}%
        {\normalsize\bfseries Proton Agent:}%
        \par\smallskip
        \tiny
      }
   ]}
  {\end{tcolorbox}\medskip}

\title{\textbf{A Tool-Grounded AI Agent for Gauge Coupling Unification and Proton Decay}}
\author{Shaikh Saad}
\affil{Jo\v{z}ef Stefan Institute, Jamova 39,  Ljubljana, Slovenia}
\affil{\texttt{\color{blue_munsell}shaikh.saad@ijs.si}}
\date{}

\begin{document}
\maketitle

%%%%%%%%%%%%%%%%%%%%%%%%%%%%%%%%%%%%%%%%%%%%%%%%%%%%%%
\begin{abstract}
We present \code{proton\_agent}, a tool-grounded AI agent that enables
natural-language interaction with computational tools for gauge coupling
unification and proton decay in grand unified theories.
The agent translates user requests into computational tasks, selects
and executes the appropriate tools, and interprets their outputs.
All numerical calculations, model evaluations, experimental comparisons,
parameter scans, and visualizations are handled by a deterministic
scientific backend to ensure numerical accuracy and reproducibility. A central objective of this framework is to enable fully local deployment using free and open-source LLMs, while retaining compatibility with commercial LLM APIs.

The framework computes one- and two-loop gauge beta coefficients
for user-defined scalar and fermion multiplets and performs gauge
coupling evolution through the Standard Model and various intermediate
symmetry-breaking stages of $SO(10)$, including
$SU(3)_C \times SU(2)_L \times SU(2)_R \times U(1)_{B-L}$,
$SU(3)_C \times SU(2)_L \times U(1)_R \times U(1)_{B-L}$,
$SU(4)_C \times SU(2)_L \times U(1)_R$, and
$SU(4)_C \times SU(2)_L \times SU(2)_R$.
For proton decay, the framework evaluates gauge-mediated dimension-six
contributions in non-supersymmetric $SU(5)$ and $SO(10)$, as well as
dressed dimension-five contributions in their supersymmetric counterparts.
It supports benchmark calculations for user-specified
parameter points, random and grid-based parameter scans,
Markov-chain Monte Carlo exploration, comparisons with experimental
limits, and publication-ready two- and three-dimensional visualizations.

We describe the agent architecture, demonstrate its capabilities
through representative benchmark calculations and regression tests,
and discuss its current physics scope.
The code is publicly available at  \href{https://github.com/saad-hep/proton_agent}{\faGithub~proton\_agent}. 
\end{abstract}
%%%%%%%%%%%%%%%%%%%%%%%%%%%%%%%%%%%%%%%%%%%%%%%%%%%%%%

\tableofcontents
%%%%%%%%%%%%%%%%%%%%%%%%%%%%%%%%%%%%%%%%%%%%%%%%%%%%%%
%%%%%%%%%%%%%%%%%%%%%%%%%%%%%%%%%%%%%%%%%%%%%%%%%%%%%%
\section{Introduction}
\label{sec:introduction}

Large language models (LLMs)~\cite{zhao2026survey} have transformed
the way users interact with computers, allowing them to ask
questions and receive responses in natural language.
However, the responses generated by LLMs are statistical in
nature and are not guaranteed to be correct.
While this may be acceptable for many everyday applications,
scientific research demands a much higher level of accuracy,
reliability, and reproducibility.
In particular, numerical calculations and theoretical predictions
must be based on well-established mathematical methods and
validated computational tools, rather than relying on the
language model itself.
This has motivated the development of scientific artificial
intelligence (AI) agents that combine the natural-language
capabilities of LLMs with dedicated computational tools.
The development of such agents is still in its early stages,
opening up a promising research direction toward building
increasingly reliable, efficient, and autonomous frameworks
for scientific computation.

In high-energy physics, recent community white papers have
outlined a vision for AI-enabled research in both experimental
and theoretical particle physics
~\cite{Aarrestad:2026xrs,CraigGroup:2026wof}.
Agentic AI systems have already demonstrated
the ability to support simulation workflows, automate event
generation, reproduce published analyses, and assist both
novice and expert researchers
~\cite{Menzo:2025cim,plehn2026madagents,moreno2026autonomousphysics,
colliderbench2026,Esmail:2026jpb,Qiu:2026iby,Agrawal:2026lvg,
Menzo:2026qrl,Alexander:2026lpw,Faroughy:2026dkj,
Desai:2026nmx,Lucente:2026kgh,Wang:2026jjn,Saad:2026pan,
Hammad:2026ged,Diefenbacher:2026azr,Guo:2026ifi,
Hammad:2026bvw,Menzo:2026xat,Kumar:2026zpo,Menzo:2026eet}.
These developments demonstrate the growing potential
of AI agents to move beyond conventional question-answering
systems toward more sophisticated scientific workflows,
reducing technical barriers and facilitating complex
computational tasks.

In this work, we introduce \PD, a \emph{domain-specific}
scientific AI agent designed to facilitate computational studies
of Grand Unified Theories (GUTs)~\cite{Pati:1974yy, Georgi:1974sy, Georgi:1974yf, Georgi:1974my, Fritzsch:1974nn}, with a particular focus on gauge
coupling unification and proton decay. Proton decay is among the
most distinctive experimental signatures of grand unification,
providing one of the strongest probes of these theories.
The unification scale and the corresponding unified gauge coupling
play a crucial role in determining the proton lifetime, particularly
for gauge-mediated decay.
To facilitate such calculations, \PD accepts user-defined scalar
and fermion multiplets through natural-language instructions,
computes the corresponding one- and two-loop gauge beta coefficients,
and performs gauge coupling evolution through the Standard Model (SM),
the Minimal Supersymmetric Standard Model (MSSM), and various
intermediate symmetry-breaking stages of $SO(10)$, including
left-right (LR) symmetry and Pati-Salam (PS) groups.

Beyond gauge coupling evolution, \PD provides
a natural-language interface to dedicated scientific tools for
computing gauge-mediated dimension-six proton decay in
non-SUSY $SU(5)$ and $SO(10)$, as well as dressed
dimension-five contributions in their supersymmetric counterparts.
Realistic proton decay predictions involve several interconnected
calculations, including fermion flavor rotations, relevant Wilson coefficients, plugging the right hadronic matrix elements, and comparisons with experimental limits
for individual decay channels. Exploring these predictions over
a high-dimensional parameter space further adds to the computational
complexity. To streamline this process, the agent supports
calculations at user-specified benchmark points, random and grid-based
parameter scans, Markov-chain Monte Carlo exploration, automated
comparisons with experimental constraints, and publication-ready
two- and three-dimensional visualizations.

The primary objective of \PD is to assist researchers in
performing accurate and reproducible proton decay calculations
while facilitating gauge coupling unification studies through
a unified, natural-language-driven computational framework. Moreover, a central objective of this framework is to enable fully local deployment using free and open-source LLMs, without requiring proprietary model APIs, while retaining compatibility with commercial LLM services.

The paper is organized as follows. Section~\ref{sec:physics} describes
the implemented gauge-coupling and proton-decay calculations. Section~\ref{sec:architecture} presents the agent architecture,
execution flow, and consistency checks.
Section~\ref{sec:examples} illustrates representative calculations
and analyses through natural-language requests.
We conclude in Sec.~\ref{sec:conclusion}. Furthermore, Appendix~\ref{app:interface_reference} summarizes the workflows, inputs,
and conventions.

%%%%%%%%%%%%%%%%%%%%%%%%%%%%%%%%%%%%%%%%%%%%%%%%%%%%%%
%%%%%%%%%%%%%%%%%%%%%%%%%%%%%%%%%%%%%%%%%%%%%%%%%%%%%%
\section{Physics scope}
\label{sec:physics}

%%%%%%%%%%%%%%%%%%%%%%%%%%%%%%%%%%%%%%%%%%%%%%%%%%%%%%
%%%%%%%%%%%%%%%%%%%%%%%%%%%%%%%%%%%%%%%%%%%%%%%%%%%%%%
\subsection{Gauge-coupling evolution}

The evolution of gauge couplings, $g_i$, is governed by the
renormalization group equations (RGEs), which we implement
up to two-loop order using the convention~\cite{Machacek:1983tz}
\begin{equation}
  \mu\frac{d g_i}{d\mu}
  = \frac{g_i^3}{(4\pi)^2}b_i
  + \frac{g_i^3}{(4\pi)^4}\sum_j B_{ij}g_j^2,
  \label{eq:rge}
\end{equation}
where $b_i$ and $B_{ij}$ denote the one- and two-loop
beta-function coefficients, respectively. 
For the SM, the one-loop coefficients are
$b_i^{\rm SM}=(41/10,-19/6,-7)$~\cite{Georgi:1974yf,Fischler:1981is,Jones:1981we}, where the hypercharge
coupling follows the conventional $SU(5)$ normalization. At two-loop order, we retain only the gauge contributions to the beta functions, neglecting Yukawa-dependent terms\footnote{
Third-generation Yukawa couplings are evolved at one loop and included in the two-loop gauge RGEs only in the standalone SM workflow; any other gauge group and MSSM scenarios use gauge-only evolution throughout.
}. 
The framework accepts additional scalar and fermion multiplets
subject to the representation and interface restrictions summarized
in Table~\ref{tab:beta_representations}. Their gauge contributions
to the beta functions are evaluated using Dynkin indices,
quadratic Casimir invariants, and multiplicities. Scalars are treated as complex and fermions as Weyl by default;
explicitly specified real scalars and Dirac fermions receive
the corresponding counting factors.

The RGE backend supports gauge coupling evolution within
both the SM and the MSSM, as well as through several intermediate
symmetry-breaking stages of $SO(10)$~\cite{Slansky:1981yr}:
\begin{align}
G_{3221}&=SU(3)_C\times SU(2)_L\times SU(2)_R\times U(1)_{B-L},\\
G_{3211}&=SU(3)_C\times SU(2)_L\times U(1)_R\times U(1)_{B-L},\\
G_{422}&=SU(4)_C\times SU(2)_L\times SU(2)_R,\\
G_{421}&=SU(4)_C\times SU(2)_L\times U(1)_R.
\end{align}
Particle mass thresholds are incorporated through piecewise RGE
evolution, with the beta-function coefficients updated as
the corresponding thresholds are crossed. At each symmetry-breaking
scale, the gauge couplings are matched according to the
appropriate group-theoretical relations. Gauge kinetic mixing~\cite{Holdom:1985ag,delAguila:1988jz}
between the two $U(1)$ factors in $G_{3211}$ is neglected.
For transitions from the SM to the MSSM, the framework
performs the appropriate conversion~\cite{Martin:1993yx} of gauge couplings
from the $\overline{\rm MS}$ to the $\overline{\rm DR}$
renormalization scheme at the user-specified supersymmetry
scale before continuing the evolution.

For theories in which the gauge symmetry remains that of
the SM below the unification scale, as commonly considered
in $SU(5)$ GUTs, a useful one-loop diagnostic can be
constructed from the differences between the gauge
beta-function coefficients:
\begin{equation}
 \mathcal{S}_{ij}=(b_i-b_j)_{\rm SM}
 +\sum_J(\Delta b_i^{(J)}-\Delta b_j^{(J)})r_J,
 \qquad
 r_J=\frac{\ln(M_U/M_J)}{\ln(M_U/M_Z)},
 \label{eq:thresholdweight}
\end{equation}
where $M_U$ is the unification scale, $M_J$ denotes the
mass of an additional multiplet, and $r_J$ represents
the fraction of the logarithmic energy interval over which
the multiplet contributes to the running.
The one-loop unification conditions can then be expressed as~\cite{Giveon:1991zm}  
\begin{equation}
 \frac{\mathcal{S}_{23}}{\mathcal{S}_{12}}
 =\frac{\alpha_2^{-1}(M_Z)-\alpha_3^{-1}(M_Z)}
 {\alpha_1^{-1}(M_Z)-\alpha_2^{-1}(M_Z)},
 \qquad
 \ln\frac{M_U}{M_Z}
 =\frac{2\pi[\alpha_1^{-1}(M_Z)-\alpha_2^{-1}(M_Z)]}
 {\mathcal{S}_{12}}.
 \label{eq:unification}
\end{equation}
For a given threshold weight $r_J$, the corresponding
multiplet mass can be determined from
\begin{equation}
M_J=M_Z\left(\frac{M_U}{M_Z}\right)^{1-r_J}.
\end{equation}
The dedicated unification tool evaluates these conditions
and performs the threshold inversion, allowing \PD
to assess how individual multiplets affect gauge coupling
unification and identify the mass scales required to
achieve it.

Further details on the gauge-coupling evolution and unification inputs and run-options are relegated to Appendix~\ref{app:interface_reference}.

\begin{table}[t!]
\centering
\small
\renewcommand{\arraystretch}{1.15}
\setlength{\tabcolsep}{5pt}
\begin{tabularx}{\textwidth}{
>{\raggedright\arraybackslash}p{0.17\textwidth}
>{\raggedright\arraybackslash}X
>{\raggedright\arraybackslash}p{0.28\textwidth}}
\toprule
\textbf{Group or property} & \textbf{Notation in request} &
\textbf{Current scope or restriction}\\

\midrule
$SU(2)_{L,R}$ &
State the integer dimension $d=2j+1$ in the appropriate tuple position. &
Any positive integer dimension is accepted.\\

\midrule
$SU(3)_C$ &
Use the named labels
$\mathbf{1},\mathbf{3},\overline{\mathbf{3}},\mathbf{6},
\overline{\mathbf{6}},\mathbf{8},\mathbf{10},\overline{\mathbf{10}},
\mathbf{15},\overline{\mathbf{15}},\mathbf{15}'$.
Plain-text conjugates can be written as \code{bar3}, \code{bar6},
\code{bar10}, \code{bar15}; use \code{p15} for $\mathbf{15}'$. &
This is a finite label set.
\\

\midrule
$SU(4)_C$ &
Use $\mathbf{1},\mathbf{4},\overline{\mathbf{4}},\mathbf{6},
\mathbf{10},\overline{\mathbf{10}},\mathbf{15},\mathbf{20},\mathbf{20}'$;
plain-text labels include \code{bar4}, \code{bar10}, and \code{p20}. &
Finite lookup only. 
The labels \code{20} and \code{p20} distinguish the two listed dimension-20 irreps.\\

\midrule
$G_{321}$ (SM/MSSM) &
State $(R_3,d_L,Y)$, for example,
``a complex scalar \code{(6,2,-1/6)}.'' &
Hypercharge may be given as an exact fraction. \\
$G_{3221}$ (LR) &
State $(R_3,d_L,d_R,B-L)$. &
The final charge is $B-L$.\\
$G_{3211}$ &
State $(R_3,d_L,R,(B-L)/2)$. &
The final charge is $(B-L)/2$.\\
$G_{422}$ (PS) &
State $(R_4,d_L,d_R)$. &
Use the finite $SU(4)$ labels above.\\
$G_{421}$ &
State $(R_4,d_L,R)$. &
The final entry is the $U(1)_R$ charge.\\
\bottomrule
\end{tabularx}
\caption{Multiplet notation for RGE requests.}
\label{tab:beta_representations}
\end{table}

%%%%%%%%%%%%%%%%%%%%%%%%%%%%%%%%%%%%%%%%%%%%%%%%%%%%%%
%%%%%%%%%%%%%%%%%%%%%%%%%%%%%%%%%%%%%%%%%%%%%%%%%%%%%%
\subsection{Non-supersymmetric proton decay}\label{sec:pd}

The non-supersymmetric $SU(5)$ and $SO(10)$ modules describe
gauge-boson-mediated dimension-six proton decay~\cite{Pati:1973rp}.
Schematically, the partial proton decay width can be written as
\begin{equation}
 \Gamma(p\to M\ell)\sim \frac{g_{\rm GUT}^4}{M_X^4}
 A^2|\mathcal{M}_{\rm had}|^2|\mathcal{F}_{\rm flavor}|^2\Phi,
 \label{eq:d6}
\end{equation}
where $M_X$ denotes the mass of the heavy gauge boson
mediating proton decay, $A$ collects the renormalization
factors, $\mathcal{M}_{\rm had}$ represents the hadronic
matrix element, $\mathcal{F}_{\rm flavor}$ encodes the
dependence on fermion flavor rotations, and $\Phi$ denotes
the phase-space factor. The hadronic matrix elements
are taken from lattice-QCD calculations
\cite{Aoki:2017puj,Yoo:2021gql}.

For the gauge-mediated proton decay calculations in $SO(10)$,
we follow the formalism summarized in Appendix~D of
Ref.~\cite{Saad:2026ucf} (which adopted computations from Refs.~\cite{FileviezPerez:2004hn,Nath:2006ut}), adopting the flavor-rotation conventions
defined in Eqs.~(D.14)-(D.15) therein. The same expressions apply to
$SU(5)$ upon setting $K=0$ (the quantity, $K$, is defined right above Eq. (D.10) of Ref.~\cite{Saad:2026ucf}), with the corresponding flavor
rotations following the conventions of
Ref.~\cite{Dorsner:2024seb}. The two GUT implementations use different conventions for
the flavor rotations. The $SO(10)$ module accepts the full
set of unitary matrices $U_{L,R}$, $D_{L,R}$, $E_{L,R}$,
and $N_L$, whereas the $SU(5)$ module follows the notation adopted  in Eqs.~(8)-(11) of Ref.~\cite{Dorsner:2024seb} (see  Table~\ref{tab:proton_inputs}). The appropriate mapping between these conventions are
\begin{equation}
U_L=U,\quad  D_L=D,\quad  E_L=E,\quad  N_L= N,\quad    U_R=U_c^*,\quad D_R=D_c^*,\quad E_R=E_c^*.
\end{equation}
The agent preserves these distinctions when interpreting user
inputs and selecting the appropriate computational tools.
For either GUT, proton decay lifetimes can be evaluated
directly from user-specified values of $(M_{\rm GUT},g_{\rm GUT})$
and flavor-rotation matrices, or from benchmark input files.

The implementation parameterizes a general $3\times3$ unitary
flavor rotation as
\begin{align}
\mathcal{U}&=D_1\,V(\theta_{12},\theta_{23},\theta_{13},\delta)\,D_2,
\label{eq:general_unitary}\\
D_1&=\operatorname{diag}(e^{i\varphi_1},e^{i\varphi_2},e^{i\varphi_3}),
\qquad
D_2=\operatorname{diag}(e^{i\varphi_4},e^{i\varphi_5},1),
\notag
\end{align}
where
\begin{equation}
V=\begin{pmatrix}
c_{12}c_{13} & s_{12}c_{13} & s_{13}e^{-i\delta}\\
-s_{12}c_{23}-c_{12}s_{23}s_{13}e^{i\delta}
& c_{12}c_{23}-s_{12}s_{23}s_{13}e^{i\delta} & s_{23}c_{13}\\
s_{12}s_{23}-c_{12}c_{23}s_{13}e^{i\delta}
& -c_{12}s_{23}-s_{12}c_{23}s_{13}e^{i\delta} & c_{23}c_{13}
\end{pmatrix},
\label{eq:ckm_like_rotation}
\end{equation}
with $s_{ij}=\sin\theta_{ij}$ and $c_{ij}=\cos\theta_{ij}$.
Each rotation has nine real parameters: three mixing angles
$(\theta_{12},\theta_{23},\theta_{13})$ and six phases
$(\delta,\varphi_1,\ldots,\varphi_5)$.
Setting the final diagonal entry of $D_2$ to unity removes the
redundancy between the two diagonal phase matrices without restricting
the general unitary rotation. The angle and phase ranges used for
unrestricted sampling are $[0,\pi/2]$ and $[0,2\pi)$, respectively.

%%%%%%%%%%%%%%%%%%%%%%%%%%%%%%%%%%%%%%%%%%%%%%%%%%%%%%
%%%%%%%%%%%%%%%%%%%%%%%%%%%%%%%%%%%%%%%%%%%%%%%%%%%%%%
\subsection{Supersymmetric proton decay}

In supersymmetric GUTs, proton decay mediated by color-triplet
Higgs fields through dimension-five operators~\cite{Sakai:1981pk,Weinberg:1981wj} is particularly
important, as these contributions can lead to proton lifetimes
in tension with experimental limits. Integrating out the heavy
color-triplet Higgs fields generates baryon-number-violating
dimension-five operators in the effective superpotential.
These operators are subsequently dressed by loop diagrams
involving superpartners, particularly winos and higgsinos,
which convert them into effective four-fermion operators
responsible for proton decay.
At the amplitude level, the dependence on the color-triplet
mass can be expressed schematically as
\begin{equation}
 \mathcal{A}_{d=5}\propto \frac{1}{M_T},\qquad
 \Gamma\propto M_T^{-2},\qquad
 \tau\propto M_T^2,
 \label{eq:d5}
\end{equation}
when all other parameters are held fixed.
Here $M_T$ denotes the color-triplet Higgs mass,
which is distinct from the masses of the heavy
$X^{(\prime)}$ and $Y^{(\prime)}$ gauge bosons
mediating dimension-six proton decay.

\begin{table}[t!]
\centering

\renewcommand{\arraystretch}{1}
\setlength{\tabcolsep}{5pt}
\small

\begin{tabularx}{\textwidth}{
>{\raggedright\arraybackslash}p{0.17\textwidth}
>{\raggedright\arraybackslash}p{0.20\textwidth}
>{\raggedright\arraybackslash}X
}
\hline\hline
\textbf{GUT framework}
&
\textbf{Proton decay mechanism}
&
\textbf{Input parameters}
\\
\hline

Non-SUSY $SU(5)$
&
Gauge-mediated dimension-six (implementation from Ref.~\cite{Saad:2026ucf})
&
Unification scale $M_{\rm GUT}$,
unified gauge coupling $g_{\rm GUT}$,
and the full $3\times3$ flavor rotation matrices
$U$, $D$, $E$, $U_c$, $D_c$, $E_c$, and $N$.  
\par\smallskip
Alternatively, 
the reduced set
$U,E,U_c,D_c,E_c$, five external CKM phases,
and six PMNS phases (one Dirac and five external).
The CKM angles and Dirac phase and the PMNS angles are fixed
to the implemented benchmark values Eq.~\eqref{eq:reduced_flavor_mixing_inputs}; 
$D=UU_{\rm CKM}$ and $N=EU_{\rm PMNS}$ are derived.
\\

\hline

Non-SUSY $SO(10)$
&
Gauge-mediated dimension-six (implementation from Ref.~\cite{Saad:2026ucf})
&
Unification scale $M_{\rm GUT}$,
unified gauge coupling $g_{\rm GUT}$,
and $3\times3$ flavor rotation matrices
$U_L$, $U_R$, $D_L$, $D_R$,
$E_L$, $E_R$, and $N_L$.
\\

\hline

SUSY $SU(5)$
&
Color-triplet-mediated dimension-five,
including gluino, chargino, and neutralino dressing (implementation from Ref.~\cite{Goto:1998qg})
&
Universal scalar mass $m_0$,
dimensionless trilinear coefficient $A_X$,
$\tan\beta$,
color-triplet mass $M_T$,
higgsino mass parameter $\mu$,
a common gaugino-mass parameter $M_{1/2}$,
with optional independent inputs $M_1$, $M_2$, and $M_3$,
$3\times3$ Yukawa matrices $f_U$, $f_D$, and $f_E$,
and flavor phases $\phi_{13}$ and $\phi_{23}$ defining
$P=\operatorname{diag}(e^{i\phi_{13}},e^{i\phi_{23}},1)$;
file-only dressing gauge couplings $g_2$ and $g_3$.
Unless specified independently, the gaugino masses follow
$(M_1,M_2,M_3)=(1,2,6)M_{1/2}$.
Each explicitly supplied $M_i$ overrides its corresponding
default relation; the remaining masses retain their
$M_{1/2}$ dependence. 
Omitted benchmark inputs use the implemented defaults, for example, the Yukawa values as in Eqs.~\eqref{eq:fU}-\eqref{eq:fE}.
\\

\hline

SUSY $SO(10)$
&
Color-triplet-mediated dimension-five,
including wino and higgsino dressing (implementation from Ref.~\cite{Severson:2015dta,Severson:2016rqb,Mohapatra:2018biy})
&
$3\times3$ Yukawa matrices $h$, $f$, and $g$
associated with $10_H$, $\overline{126}_H$,
and $120_H$, respectively;
$\tan\beta$;
Higgs-doublet mixing parameters
$(\theta_u,\phi_u,\psi_u)$ and
$(\theta_d,\phi_d,\psi_d)$ (as defined in Eqs.~\eqref{eq:so10last}-\eqref{eq:so10first});
seesaw scales $v_L$ and $v_R$;
color-triplet mixing coefficients $x_0,\ldots,x_{10}$
and $y_5,y_7,y_9,y_{10}$;
color-triplet mass $M_T$, wino mass $M_{\widetilde W}$,
higgsino mass parameter $\mu$, and common sfermion mass
$m_{\widetilde q}$.
Omitted benchmark inputs use the implemented defaults.
\\

\hline\hline
\end{tabularx}
\caption{Summary of the supported proton decay frameworks    and their physical input parameters. }
\label{tab:proton_inputs}
\end{table}

For supersymmetric $SU(5)$, the backend follows the formalism of Ref.~\cite{Goto:1998qg}, incorporating gluino, chargino, and neutralino dressing contributions to both $LLLL$ and $RRRR$ dimension-five operators. The calculation accounts
for the relevant flavor-dependent coefficients and
their dependence on the supersymmetric particle spectrum.
The user-specified inputs (see  Table~\ref{tab:proton_inputs}) include the universal scalar
mass $m_0$, the trilinear coupling $A_X$, $\tan\beta$,
the color-triplet Higgs mass $M_T$, the higgsino mass
parameter $\mu$, the universal gaugino mass $M_{1/2}$,
the Yukawa matrices $f_U$, $f_D$, and $f_E$,
and two independent flavor phases $\phi_{13}$ and $\phi_{23}$.
The benchmark file can also specify the dressing gauge couplings
$g_2$ and $g_3$; the three gaugino masses are derived from
$M_{1/2}$ as listed in Table~\ref{tab:proton_inputs}.

For supersymmetric $SO(10)$, the implementation follows
the proton decay formalism developed in
Refs.~\cite{Severson:2015dta,Severson:2016rqb,Mohapatra:2018biy} (see also, Refs.~\cite{Babu:1998wi,Babu:2018tfi}),
incorporating both $LLLL$ and $RRRR$ dimension-five operators
and their corresponding dressing contributions.
In this framework, the Yukawa coupling matrices
$h$, $f$, and $g$, associated with the Higgs
representations $10_H$, $\overline{126}_H$, and $120_H$,
respectively, determine the flavor structure of
the dimension-five operators.
Additional inputs include $\tan\beta$, the six Higgs-doublet
mixing parameters, the seesaw scales $v_L$ and $v_R$,
the color-triplet mixing coefficients listed in
Table~\ref{tab:proton_inputs},
and the masses of the color-triplet Higgs fields,
winos, higgsinos, and sfermions.
These parameters enter the calculation of the
effective operators and their dressing contributions,
which ultimately determine the proton decay amplitudes
and partial lifetimes.

Since proton decay calculations require the original Yukawa couplings rather than the rescaled ones commonly used in fermion mass fits, we clarify below the conversion conventions adopted in our analysis.
The electroweak VEVs are conveniently parametrized as
\begin{equation}
v_u=v\sin\beta,\qquad
v_d=v\cos\beta,\qquad
v=174~\mathrm{GeV},
\label{eq:so10first}
\end{equation}
with the individual contributions expressed through
hyperspherical coordinates:
\begin{equation}
\begin{pmatrix}
v_{10}^a\\
v_{126}^a\\
v_{120_1}^a\\
v_{120_{15}}^a
\end{pmatrix}
=
v_a
\begin{pmatrix}
\cos\theta_a\\
\sin\theta_a\cos\phi_a\\
\sin\theta_a\sin\phi_a\cos\psi_a\\
\sin\theta_a\sin\phi_a\sin\psi_a
\end{pmatrix},
\qquad a=u,d.
\end{equation}
In the fermion mass convention $f^c M f$, we choose a
flavor basis in which $h$ is real, diagonal, and positive,
while $f$ and $g$ are complex symmetric and antisymmetric,
respectively. The fermion mass matrices are then given by
\begin{align}
M_U &= v_{10}^u h + v_{126}^u f
     + (v_{120_1}^u+v_{120_{15}}^u)g,\\
M_D &= v_{10}^d h + v_{126}^d f
     + (v_{120_1}^d+v_{120_{15}}^d)g,\\
M_E &= v_{10}^d h -3v_{126}^d f
     + (v_{120_1}^d-3v_{120_{15}}^d)g,\\
M_{\nu_D} &= v_{10}^u h -3v_{126}^u f
     + (v_{120_1}^u-3v_{120_{15}}^u)g,\\
M_N &= v_L f-M_{\nu_D}^T(v_R f)^{-1}M_{\nu_D}.
\label{eq:so10last}
\end{align}
Here, $v_L$ and $v_R$ denote the VEVs associated with
the left- and right-handed Majorana neutrino masses,
respectively, with $M_N$ receiving both type-I~\cite{Minkowski:1977sc,Gell-Mann:1979vob,Yanagida:1979as,Schechter:1980gr,Glashow:1979nm,Mohapatra:1979ia} and
type-II~\cite{Magg:1980ut,Schechter:1980gr,Lazarides:1980nt,Mohapatra:1980yp} seesaw contributions.

Further details on the proton decay inputs and run-options are relegated to Appendix~\ref{app:interface_reference}.

%%%%%%%%%%%%%%%%%%%%%%%%%%%%%%%%%%%%%%%%%%%%%%%%%%%%%%
%%%%%%%%%%%%%%%%%%%%%%%%%%%%%%%%%%%%%%%%%%%%%%%%%%%%%%
\subsection{Experimental comparison}\label{sec:exp}

For each parameter point, \PD computes the partial proton
lifetimes for the relevant decay channels and compares them
with the corresponding experimental lower limits.
Table~\ref{tab:nucleon_decay} summarizes the current
experimental bounds and the projected sensitivities of
Hyper-Kamiokande after ten years of operation.

Based on these comparisons, the agent automatically classifies
each parameter point into three categories:
(i)~\textit{excluded}, if at least one predicted partial lifetime
violates an existing experimental bound;
(ii)~\textit{testable}, if the point satisfies all current
bounds but falls within the projected sensitivity of at least
one future search; and
(iii)~\textit{safe}, if the point satisfies
all current bounds and lies beyond the future sensitivities
considered here. In doing so, the agent uses the following constraint score:
\begin{align}
 \chi=   R_{\max}=\max_i\frac{\tau_i^{\rm limit}}{\tau_i^{\rm predicted}}.
\end{align}
\(R_{\max}<1\): all predicted lifetimes exceed their respective bounds.
\(R_{\max}>1\): at least one channel violates its bound.
\(R_{\max}=1\): a bound is saturated. Note, This is not a conventional statistical \(\chi^2\) and there is no summation over $i$. 
 
The agent also identifies the most constraining decay channel
and reports the corresponding experimental comparison.
This automated classification facilitates the exploration
of large parameter spaces and the identification of
phenomenologically viable and experimentally testable regions.

\begin{table}[th!]
\centering
\small
\renewcommand{\arraystretch}{1.2}
\begin{tabular}{lcc}
\toprule
\textbf{Decay channel}
& \textbf{Current bound [yr]}
& \textbf{Future sensitivity [yr]} \\
\midrule
$p\to \pi^0 e^+$
& $2.4\times 10^{34}$ \cite{Super-Kamiokande:2020wjk}
& $7.8\times 10^{34}$ \cite{Hyper-Kamiokande:2018ofw} \\

$p\to \pi^0 \mu^+$
& $1.6\times 10^{34}$ \cite{Super-Kamiokande:2020wjk}
& $7.7\times 10^{34}$ \cite{Hyper-Kamiokande:2018ofw} \\

$p\to \eta e^+$
& $1.0\times 10^{34}$ \cite{Super-Kamiokande:2017gev}
& $4.3\times 10^{34}$ \cite{Hyper-Kamiokande:2018ofw} \\

$p\to \eta \mu^+$
& $4.7\times 10^{33}$ \cite{Super-Kamiokande:2017gev}
& $4.9\times 10^{34}$ \cite{Hyper-Kamiokande:2018ofw} \\

$p\to K^0 e^+$
& $1.1\times 10^{33}$ \cite{Brock:2012ogj}
& --- \\

$p\to K^0 \mu^+$
& $3.6\times 10^{33}$ \cite{Super-Kamiokande:2022egr}
& --- \\

$p\to \pi^+ \overline{\nu}$
& $3.9\times 10^{32}$ \cite{Super-Kamiokande:2013rwg}
& --- \\

$p\to K^+ \overline{\nu}$
& $6.6\times 10^{33}$ \cite{Takhistov:2016eqm}
& $3.2\times 10^{34}$ \cite{Hyper-Kamiokande:2018ofw} \\
\bottomrule
\end{tabular}
\caption{Experimental lower limits on the partial proton
lifetimes for various decay channels, together with
the projected sensitivities of Hyper-Kamiokande
after ten years of operation.}
\label{tab:nucleon_decay}
\end{table}

%%%%%%%%%%%%%%%%%%%%%%%%%%%%%%%%%%%%%%%%%%%%%%%%%%%%%%
%%%%%%%%%%%%%%%%%%%%%%%%%%%%%%%%%%%%%%%%%%%%%%%%%%%%%%
\section{Agent architecture}
\label{sec:architecture}

\PD is designed as a freely available, domain-specific research
collaborator for human-guided calculations. The user defines
the physics question and model assumptions, while the agent
coordinates the requested calculations and presents their results.

The language model interprets requests, selects tools and their arguments,
and explains returned results. Numerical evaluation is delegated to explicit
scientific routines that can be tested independently of the model.
Within the permitted workflow, tool selection can depend on intermediate
results, allowing the agent to continue, revise, or end a multi-step task.
Selected explicit requests use direct execution paths instead of a model
choice at every step.

\subsection{Routing and execution}

\begin{figure}[H]
\centering
\resizebox{\linewidth}{!}{%
\begin{tikzpicture}[
    topbox/.style={
        draw,rounded corners,align=center,
        minimum height=8mm,text width=17mm,font=\tiny
    },
    widebox/.style={
        draw,rounded corners,align=center,
        minimum height=8mm,text width=36mm,font=\tiny
    },
    cyclebox/.style={
        draw,rounded corners,align=center,
        minimum height=8mm,text width=20mm,font=\tiny
    },
    smallbox/.style={
        draw,rounded corners,align=center,
        minimum height=8mm,text width=19mm,font=\tiny
    },
    arrow/.style={-{Latex[length=1.8mm]},thick},
    retryarrow/.style={-{Latex[length=1.6mm]},thick,dashed},
    agentframe/.style={draw=blue!65!black,ultra thick,rounded corners},
    parentframe/.style={draw,very thick,rounded corners},
    workflowframe/.style={draw=black,thin,solid,rounded corners},
    specialistframe/.style={draw=gray!75!black,thin,solid,rounded corners}
]

% Public interface and the six actual parent-StateGraph nodes.
\node[topbox,fill=gray!15,text width=14mm] (human) at (-7.45,0.45)
{Human input};

\node[topbox,fill=blue!15] (prepare) at (-5.15,0.45)
{\texttt{prepare\_turn}\\
history + model target};

\node[topbox,fill=orange!20,text width=19mm] (router) at (-2.85,0.45)
{\texttt{route\_specialist}\\
rules/history; optional\\router LLM fallback};

\node[topbox,fill=blue!15,text width=22mm] (proton) at (-0.20,1.02)
{\texttt{proton\_workflow}\\Proton specialist\\LLM + tools};

\node[topbox,fill=blue!15,text width=22mm] (rge) at (-0.20,-0.12)
{\texttt{rge\_workflow}\\RGE specialist\\LLM + tools};

\node[topbox,fill=blue!15] (audit) at (2.25,0.45)
{\texttt{audit\_artifacts}\\
record provenance only};

\node[topbox,fill=blue!15] (finalize) at (4.80,0.45)
{\texttt{finalize\_answer}\\
normalize answer/history};

\node[topbox,fill=gray!15,text width=14mm] (answer) at (7.45,0.45)
{Final answer};

\draw[arrow] (human) -- (prepare);
\draw[arrow] (prepare) -- (router);
\draw[arrow] (router) -- (proton);
\draw[arrow] (router) -- (rge);
\draw[arrow] (proton) -- (audit);
\draw[arrow] (rge) -- (audit);
\draw[arrow] (audit) -- (finalize);
\draw[arrow] (finalize) -- (answer);

% The lower panels expand the two parent workflow nodes. Each specialist
% StateGraph has exactly two nodes: agent (LLM) and tools (executor).
\node[font=\tiny\bfseries,fill=white,inner sep=0.4mm] at (-2.85,-1.15)
{Implementation of \texttt{proton\_workflow}};

\node[widebox,fill=green!15] (pdet) at (-2.85,-1.85)
{Input parsing and grounding\\explicit inputs + defaults};

\node[cyclebox,fill=violet!18,font=\fontsize{5}{6.1}\selectfont] (pllm) at (-4.25,-3.35)
{\texttt{agent}: Proton LLM\\interpret request\\choose tools\\set inputs\\read results\\write answer};

\node[cyclebox,fill=green!15,font=\fontsize{5}{6.1}\selectfont] (ptool) at (-1.30,-3.35)
{\texttt{tools}: executor\\compute physics\\check results\\save outputs};

\node[widebox,fill=red!15] (pguard) at (-2.85,-4.85)
{Result and scope checks\\numerical grounding};

\node[widebox,fill=gray!10] (presult) at (-2.85,-6.00) {Workflow result};

\draw[arrow] (pdet) -- (pllm);
\draw[arrow] ($(pllm.east)+(0,0.18)$) --
    node[above,font=\fontsize{5}{5.5}\selectfont] {calls} ($(ptool.west)+(0,0.18)$);
\draw[arrow] ($(ptool.west)+(0,-0.18)$) --
    node[below,font=\fontsize{5}{5.5}\selectfont] {results} ($(pllm.east)+(0,-0.18)$);
\draw[arrow] (pllm.south) -- ++(0,-0.25) -| (pguard.north);
\draw[arrow] (pguard) -- (presult);
\draw[retryarrow] (pguard.west) -- (-5.50,-4.85) |- (pllm.south west);
\node[font=\tiny,fill=white,inner sep=0.3mm] at (-5.19,-4.60) {retry};
% These explicitly implemented early returns bypass the specialist graph.
\draw[retryarrow] (pdet.west) -- (-5.74,-1.85) -- (-5.74,-6.00) -- (presult.west);
\node[font=\tiny,rotate=90,fill=white,inner sep=0.3mm] at (-5.74,-4.85)
{selected direct requests};

% RGE-workflow implementation.
\node[font=\tiny\bfseries,fill=white,inner sep=0.4mm] at (2.85,-1.15)
{Implementation of \texttt{rge\_workflow}};

\node[widebox,fill=green!15] (rdet) at (2.85,-1.85)
{Field/scale grounding\\explicit inputs + defaults};

\node[cyclebox,fill=violet!18,font=\fontsize{5}{6.1}\selectfont] (rllm) at (1.45,-3.35)
{\texttt{agent}: RGE LLM\\interpret request\\choose tools\\set inputs\\read results\\write answer};

\node[cyclebox,fill=green!15,font=\fontsize{5}{6.1}\selectfont] (rtool) at (4.40,-3.35)
{\texttt{tools}: executor\\compute physics\\check results\\save outputs};

\node[widebox,fill=red!15] (rguard) at (2.85,-4.85)
{Result checks\\coefficients + plots};

\node[widebox,fill=gray!10,text width=27mm] (rresult) at (1.95,-6.00)
{Workflow result};

\node[smallbox,fill=violet!18] (rinterp) at (4.65,-6.00)
{Optional LLM\\interpretation\\no tools};

\draw[arrow] (rdet) -- (rllm);
\draw[arrow] ($(rllm.east)+(0,0.18)$) --
    node[above,font=\fontsize{5}{5.5}\selectfont] {calls} ($(rtool.west)+(0,0.18)$);
\draw[arrow] ($(rtool.west)+(0,-0.18)$) --
    node[below,font=\fontsize{5}{5.5}\selectfont] {results} ($(rllm.east)+(0,-0.18)$);
\draw[arrow] (rllm.south) -- ++(0,-0.25) -| (rguard.north);
% A completed RGE plot terminates the tool loop without another LLM draft.
\draw[retryarrow] (rtool.east) -- (5.80,-3.35) -- (5.80,-4.85) -- (rguard.east);
\node[font=\tiny,rotate=90,fill=white,inner sep=0.3mm] at (5.80,-3.99) {plot complete};
\draw[arrow] (rguard) -- (rresult);
\draw[retryarrow] (rguard.west) -- (0.20,-4.85) |- (rllm.south west);
\node[font=\tiny,fill=white,inner sep=0.3mm] at (0.70,-4.60) {retry};
\draw[retryarrow] (rguard.south east) -- (rinterp.north);
\draw[arrow] (rinterp.west) -- (rresult.east);

% Matching workflow names identify these as expansions, not additional
% parent nodes. Avoid crossing the parent edges with call connectors.

\begin{scope}[on background layer]
    % Human input and final answer deliberately remain outside this frame.
    \draw[agentframe] (-6.35,2.10) rectangle (6.35,-6.80);
    \node[font=\small\bfseries,text=blue!55!black,fill=white,inner sep=1mm]
        at (0,2.10) {Proton agent: LLM-assisted workflows orchestrated by LangGraph};

    \draw[parentframe] (-6.15,1.57) rectangle (5.80,-0.80);
    \node[font=\tiny\bfseries]
        at (-4.20,1.72) {Parent StateGraph (six nodes)};

    \draw[workflowframe] (-6.00,-1.15) rectangle (-0.12,-6.60);
    \draw[workflowframe] (0.12,-1.15) rectangle (6.05,-6.60);

    \draw[specialistframe] (-5.55,-2.50) rectangle (-0.13,-4.15);
    \node[font=\tiny\bfseries,fill=white,inner sep=0.4mm]
        at (-1.65,-2.50) {StateGraph (two nodes)};

    \draw[specialistframe] (0.15,-2.50) rectangle (5.70,-4.15);
    \node[font=\tiny\bfseries,fill=white,inner sep=0.4mm]
        at (4.07,-2.50) {StateGraph (two nodes)};
\end{scope}

\end{tikzpicture}%
}

\vspace{0.5mm}

{\fontsize{5}{6}\selectfont
\setlength{\tabcolsep}{3mm}
\begin{tabular}{@{}lll@{}}
\colorbox{violet!18}{\strut\hspace{2mm}} specialist LLM
& \colorbox{orange!20}{\strut\hspace{2mm}} router / LLM fallback
& \colorbox{blue!15}{\strut\hspace{2mm}} parent LangGraph node
\\[0.7mm]
\colorbox{green!15}{\strut\hspace{2mm}} deterministic / tool execution
& \colorbox{red!15}{\strut\hspace{2mm}} checks / guardrails
& \colorbox{gray!15}{\strut\hspace{2mm}} external interface / workflow result
\end{tabular}

\vspace{0.2mm}
}
\caption{Schematic  workflow of \PD. Each specialist combines
an LLM node with a tool executor. Dashed paths indicate retries,
interpretation, or direct execution of selected requests.}
\label{fig:agent_architecture}
\end{figure}
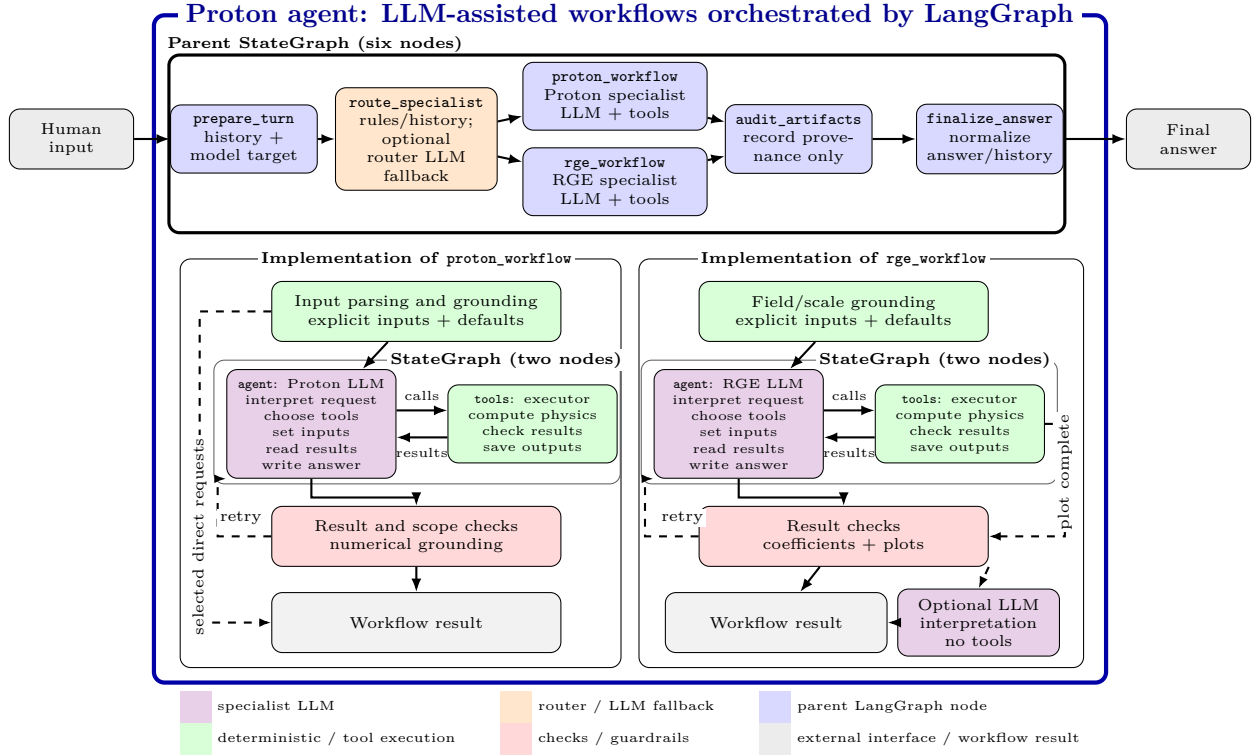

Figure~\ref{fig:agent_architecture} summarizes the execution flow and
the roles of the specialist LLMs, tool executors, and result checks.
LangGraph~\cite{LangGraph:2024} provides shared state and conditional control flow, while the
Python backend performs the scientific calculations.
The parent graph prepares the request, selects an RGE or proton-decay
specialist, records provenance, and finalizes the answer. Routing uses
request cues and recent history, with an LLM fallback when needed.
Each specialist has two nodes, \code{agent} and \code{tools}, and exposes
only its domain tools. Message state carries requests, responses, tool
calls, and observations. Conditional edges dispatch calls to the executor,
return observations to the model, or end execution. A configurable recursion limit bounds graph
execution. Compiled specialist graphs are reused by model configuration
and domain across supported providers.

\subsection{Numerical checks and workflow}

The response policy requires numerical physics claims to use tool outputs
or compatible stored results. Argument-repair and validation routines apply
documented defaults and check supported inputs. Within the tool loop,
identical repeated calls are rejected, and selected checks test parameters,
result sources, and plot files. 

Saved calculations and plots carry artifact identifiers and source
metadata; scan and MCMC samples are stored in HDF5 files. Follow-up
requests can inspect columns, compute correlations, or plot these samples
without repeating the scan. The provenance stage records tool names,
artifact identifiers, and verified plot paths found in the returned message
history.

\subsection{LLM backends and deployment}\label{sec:install}

Provider-specific adapters connect the common LangGraph workflow and
scientific tools to Ollama, remote OpenAI-compatible endpoints,
models exposed through Open WebUI, and hosted OpenAI and Anthropic
APIs. Switching between supported providers requires no changes to
the numerical backend, although tool-selection reliability and
response quality may vary across LLM models.

For local deployment, Ollama must be installed
and its service running. The package can be used directly
from the terminal or through the optional Open WebUI interface. After downloading the \PD 
repository, the user enters its root directory, activates a
Python environment with Python 3.11 or later, and installs the
package and downloads the selected model:
\begin{verbatim}
python -m pip install -e ./main
\end{verbatim}
Now, if the LLM model, for example, \code{gpt-oss:20b} is not already available in Ollama, then the user needs to download it through 
\begin{verbatim}ollama pull gpt-oss:20b\end{verbatim}
The agent server is then started with 
\begin{verbatim}
PROTON_AGENT_SERVER_API_KEY=proton-agent-local \
  proton_agent_server \
    --host 127.0.0.1 \
    --port 8001 \
    --upstream-model ollama:gpt-oss:20b
\end{verbatim}
For Open WebUI running directly on the same machine, the user
adds an OpenAI-compatible connection with base URL
\url{http://127.0.0.1:8001/v1} and API key
\code{proton-agent-local}, then selects \code{proton-agent}
in the chat interface. The server must remain running during use.
Containerized deployments require an address and server binding
that allow Open WebUI to reach the agent server.

Instead of using the Open WebUI, the user can use the \PD from terminal, the corresponding command is:
\begin{verbatim}
proton_agent --model ollama:gpt-oss:20b
\end{verbatim}

Throughout the development and testing of \PD, we used the
locally hosted \code{gpt-oss: 20b} model through Ollama.
All examples presented in this work were tested on Ubuntu 24.04,
running on a Dell Precision 5690 workstation laptop equipped with
an Intel Core Ultra 9 185H processor, 64\,GiB of system memory,
and an NVIDIA GPU with 12\,GiB of VRAM.
The model was executed with partial GPU offloading. For the
reported configuration, Ollama indicated approximately
80\% GPU and 20\% CPU model placement at a context length of
16,384 tokens. These percentages describe model placement,
rather than processor utilization, and may change with the
runtime configuration and available GPU memory.

\begin{figure}[th!]
    \centering
    \includegraphics[width=0.7\linewidth]{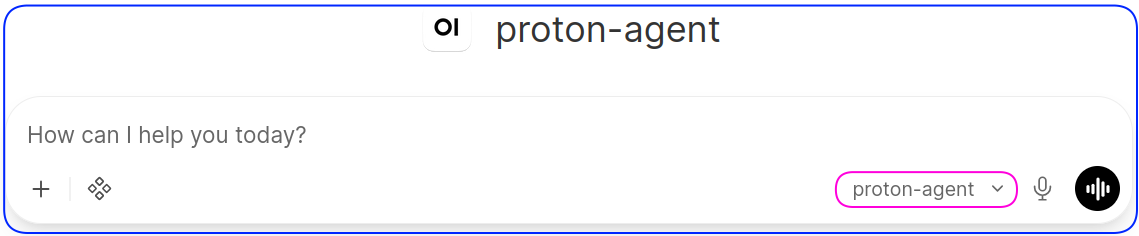}
    \caption{The Open WebUI interface with ``proton-agent'' selected.}
    \label{fig:start}
\end{figure}
%%%%%%%%%%%%%%%%%%%%%%%%%%%%%%%%%%%%%%%%%%%%%%%%%%%%%%
%%%%%%%%%%%%%%%%%%%%%%%%%%%%%%%%%%%%%%%%%%%%%%%%%%%%%%
\section{Working with the agent}
\label{sec:examples}
In this section, we demonstrate how to interact with the agent through
Open WebUI. After launching Open WebUI (see Sec.~\ref{sec:install}),
the user should select ``proton-agent'' as the model, as highlighted
by the magenta box in Fig.~\ref{fig:start}.
 
Below, we present illustrative examples demonstrating the capabilities
and usage of the agent. Numerical calculations are delegated to dedicated
scientific routines that can be tested independently of the language
model and are reproducible for fixed inputs, settings, and random seeds.
Although the agent is instructed to ground its explanations in the
computed results, LLM-generated interpretations may vary between runs
and contain errors or unsupported conclusions. Their reliability depends
on the chosen LLM and its configuration. Users should therefore
independently cross-check the physical reasoning and conclusions before
using them in scientific work.

The user submits natural-language prompts through the chat interface,
where the agent also displays its responses. Generated figures and
data files are saved in the project's output directory, and the agent
provides their file paths.

%%%%%%%%%%%%%%%%%%%%%%%%%%%%%%%%%%%%%%%%%%%%%%%%%%%%%%
%%%%%%%%%%%%%%%%%%%%%%%%%%%%%%%%%%%%%%%%%%%%%%%%%%%%%%
\subsection{Gauge-coupling evolution and unification}
\label{sec:examples:RGE}

The user can request to provide the RGE $\beta$-coefficients for a representation (see Table~\ref{tab:beta_representations}). To get both one and two loop RGE factors, the user needs to request it explicitly. For example:
\vspace{-8pt}\begin{User}
what are the new contributions to RGE beta coefficients if we add a scalar
(1,3,0) to the Standard Model; both at one and two loop
\end{User}\vspace{-23pt}
\begin{ProtonAgent}
\centering
\includegraphics[width=0.4\textwidth]{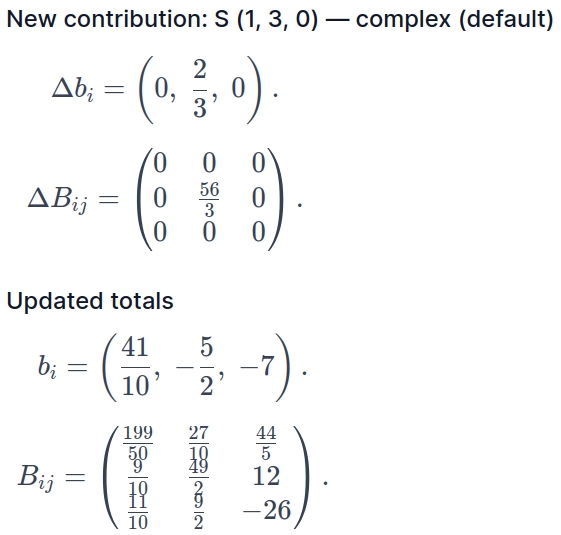}
\end{ProtonAgent}

If the user does not specify the loop order, then the agent computes only at one loop and provides the corresponding result. Moreover, if a field is \emph{real}, it needs to be specified; otherwise a \emph{complex} field is assumed by default. For example: 
\vspace{-8pt}\begin{User}
RGE beta coefficients for a real scalar (1,5,0)  
\end{User}\vspace{-23pt}
\begin{ProtonAgent}
\centering
\includegraphics[width=0.5\textwidth]{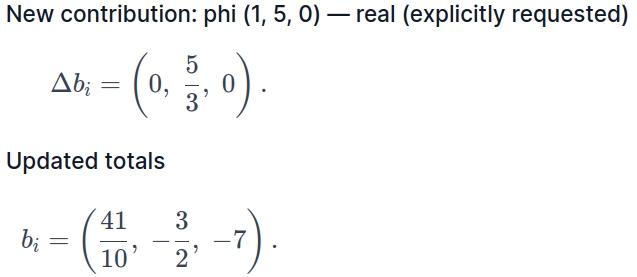}
\end{ProtonAgent}

The user can ask the agent if adding a new field to the SM can help with gauge coupling unification, and the agent will provide valuable information. For example:
\vspace{-8pt}\begin{User}
if this field, a real scalar (1,5,0), remains light, can it help with gauge coupling unification?
\end{User}\vspace{-23pt}
\begin{ProtonAgent}
\centering
\includegraphics[width=0.8\textwidth]{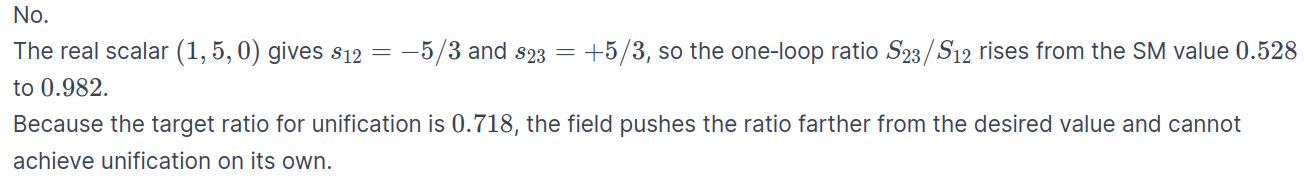}
\end{ProtonAgent}
Note, however, that in the above query, the user only mentioned `light'. Consequently, the agent sets its mass to the \(M_{Z}\) scale and provides the evaluations shown above. Given that the added field overshoots the unification criteria, the user can further ask the agent about the consequences if the field does not reside exactly at the \(M_{Z}\) scale. For example:
\vspace{-8pt}\begin{User}
what if this real scalar (1,5,0) is somewhat heavy?
\end{User}\vspace{-23pt}
\begin{ProtonAgent}
\centering
\includegraphics[width=0.8\textwidth]{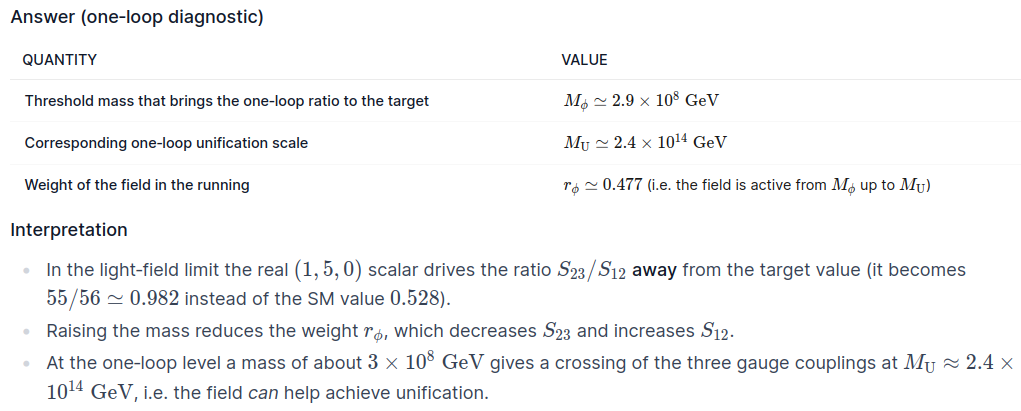}
\end{ProtonAgent}
The agent indeed checks the mass scale of the added field that best serves gauge coupling unification. If the field can achieve unification, the agent determines the required scale and reports it accordingly. The user can now ask the agent to plot this coupling unification. For example:
\vspace{-8pt}\begin{User}
plot the gauge coupling evaluations if a real scalar (1,5,0) is kept at \verb|2.9*10^8| GeV
\end{User}\vspace{-23pt}
\begin{ProtonAgent}
\centering
\includegraphics[width=0.65\textwidth]{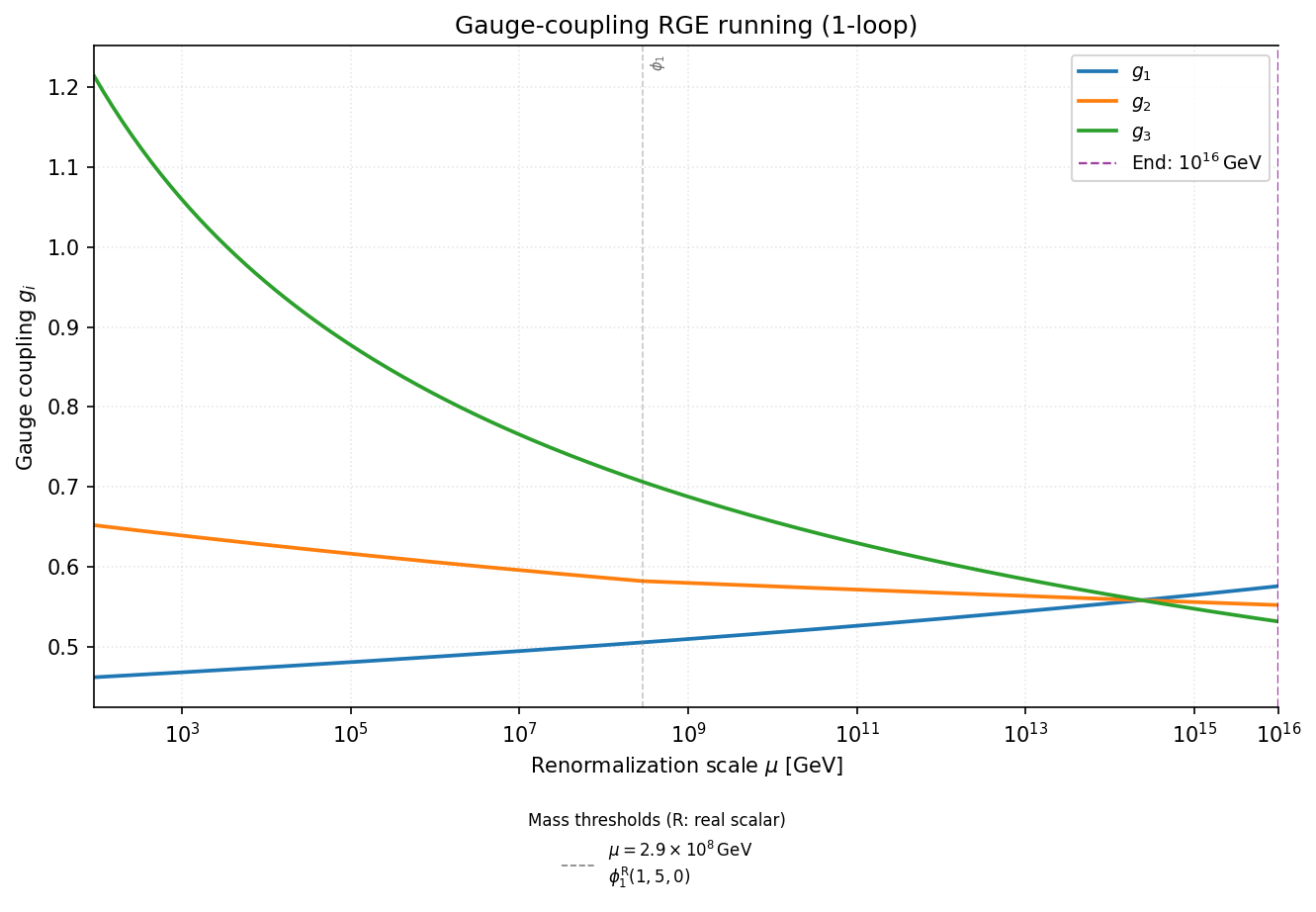}
\end{ProtonAgent}

While the previous plot shows the gauge couplings directly, it is customary to plot their inverse. The user can explicitly request this format. For example, here we consider a different multiplet than the one analyzed above:
\vspace{-8pt}\begin{User}
is a real scalar (8,3,0), sufficient to achieve gauge coupling unification; if so, what would be the needed mass of it? Keeping this derived mass fixed, plot the inverse of the gauge coupling evaluations 
\end{User}\vspace{-23pt}
\begin{ProtonAgent}
\centering
\includegraphics[width=0.8\textwidth]{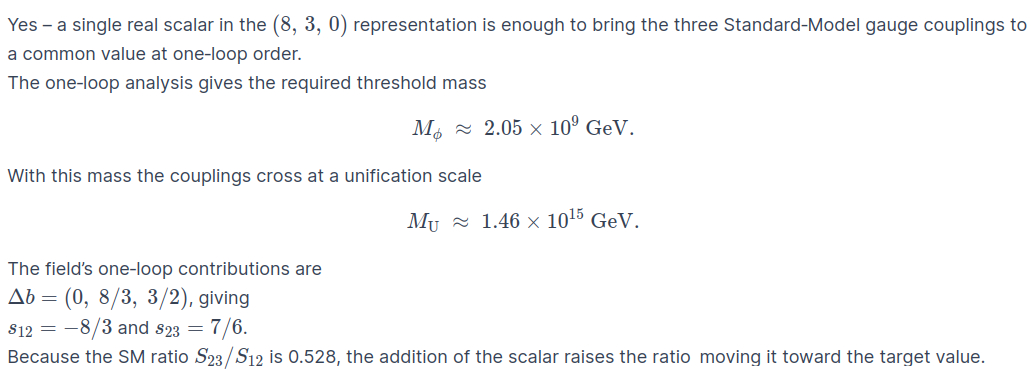}

\includegraphics[width=0.65\textwidth]{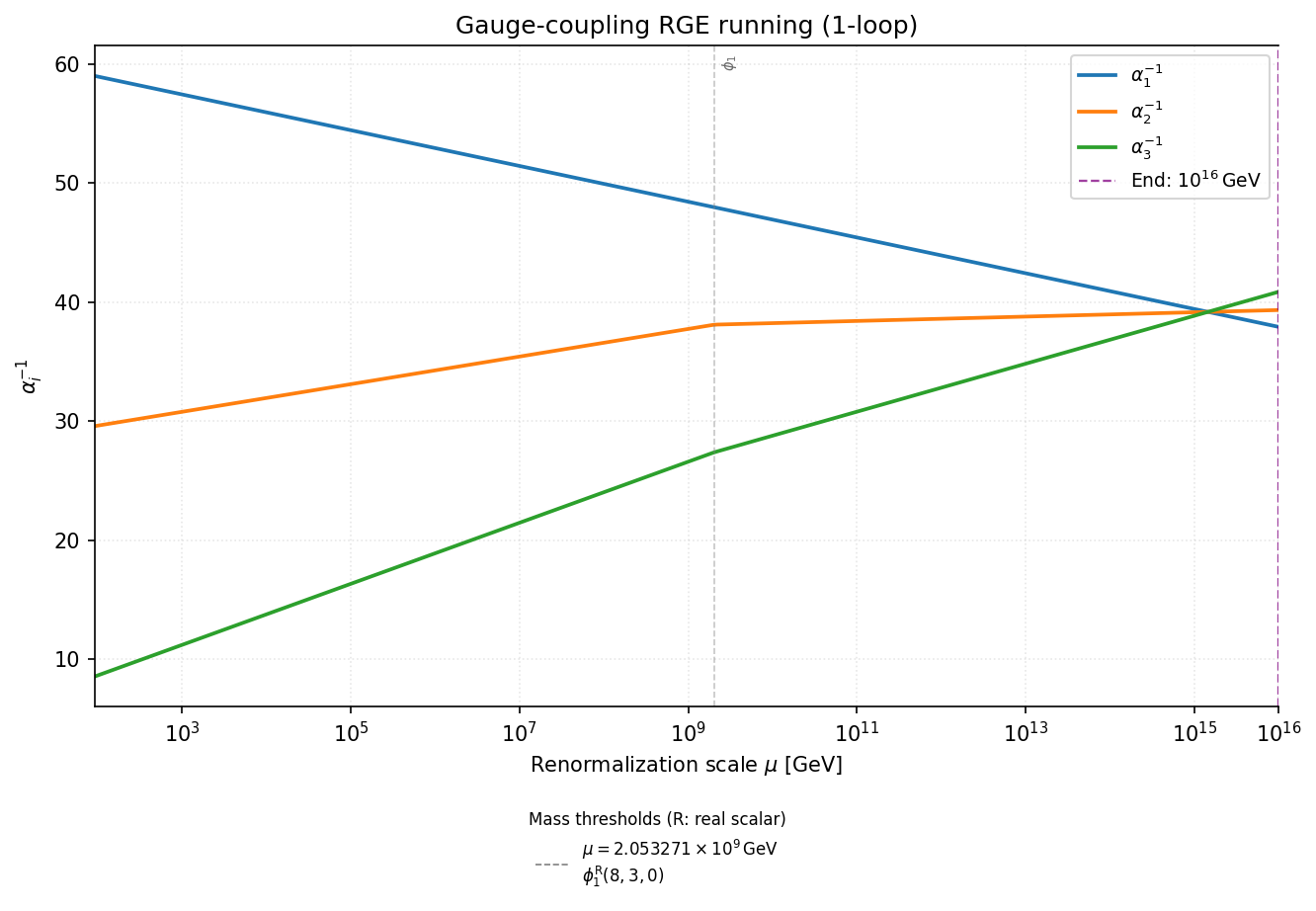}
\end{ProtonAgent}
\noindent As expected, this result exactly reproduces the recent proposal in Ref.~\cite{Dorsner:2026ltc}.

The agent can of course handle multiple fields at once. The user is expected to specify the representations of the fields as well as their respective mass scales. For example, the input might look like:
\vspace{-8pt}\begin{User}
 make a new plot with Standard Model running to 10 TeV. At 10 TeV, add four scalars: scalars (6,2,-1/6), (3,3,2/3) and  real scalars (1,3,0), (8,1,0); run them to the \verb|8.4*10^10| GeV. At \verb|8.4*10^10| GeV, add fermions (1,3,1)+(1,3,-1)+(3,2,1/6)+(3bar,2,-1/6)+(6,1,-2/3)+(6bar,1,2/3); run them to \verb|10^12| GeV. At \verb|10^12| GeV, scalars (1,4,-3/2)+(10,1,-1). Run everything upto \verb|7.7x10^15| GeV. Use one-loop RGEs. plot inverse couplings.
\end{User}\vspace{-23pt}
\begin{ProtonAgent}
\centering
\includegraphics[width=0.65\textwidth]{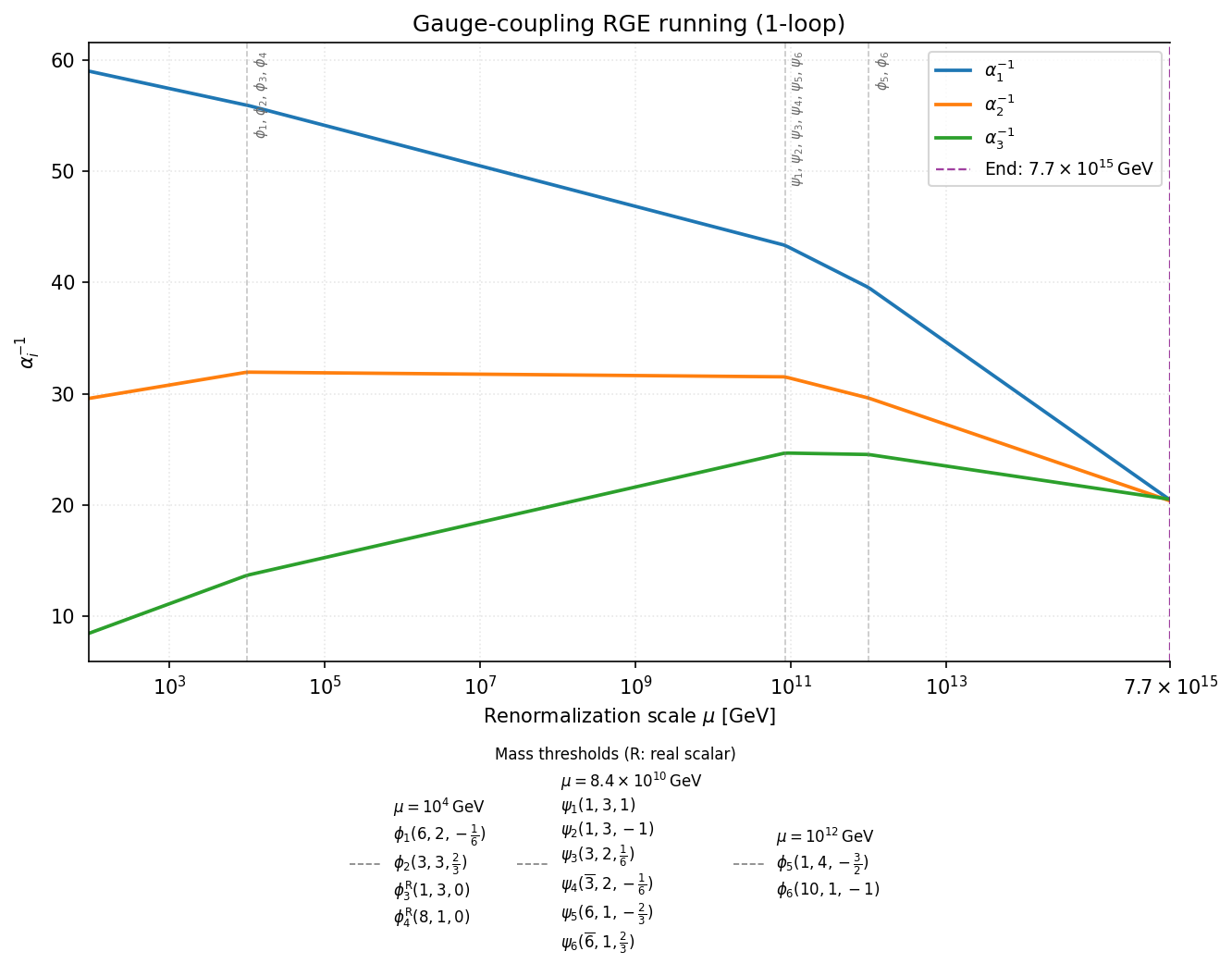}
\end{ProtonAgent}
\noindent As expected, this result exactly reproduces the proposal in Ref.~\cite{Dorsner:2019vgf}.

Furthermore, the agent can handle multiple gauge groups (see Table~\ref{tab:beta_representations}). As an example, a specific two-loop gauge coupling unification scenario within an SO(10) GUT is considered below, which reproduces the unification benchmarks presented in Ref.~\cite{Saad:2022mzu}:
\vspace{-8pt}\begin{User}
 Make a plot: run the Standard Model RGEs to 1e13 GeV, where group is changed to left-right. At 1e13 GeV, add scalars (1,1,3,-2)+(1,2,2,0); run them up to 1.62339e14 GeV. At 1.62339e14 GeV, group is changed to Pati-Salam. At 1.62339e14 GeV, add scalars (15,2,2)+(10,3,1)+(10,1,3)+(15,1,1); run them to 2.66e15 GeV. Use two-loop RGEs. plot inverse couplings.
\end{User}\vspace{-23pt}
\begin{ProtonAgent}
\centering
\includegraphics[width=0.65\textwidth]{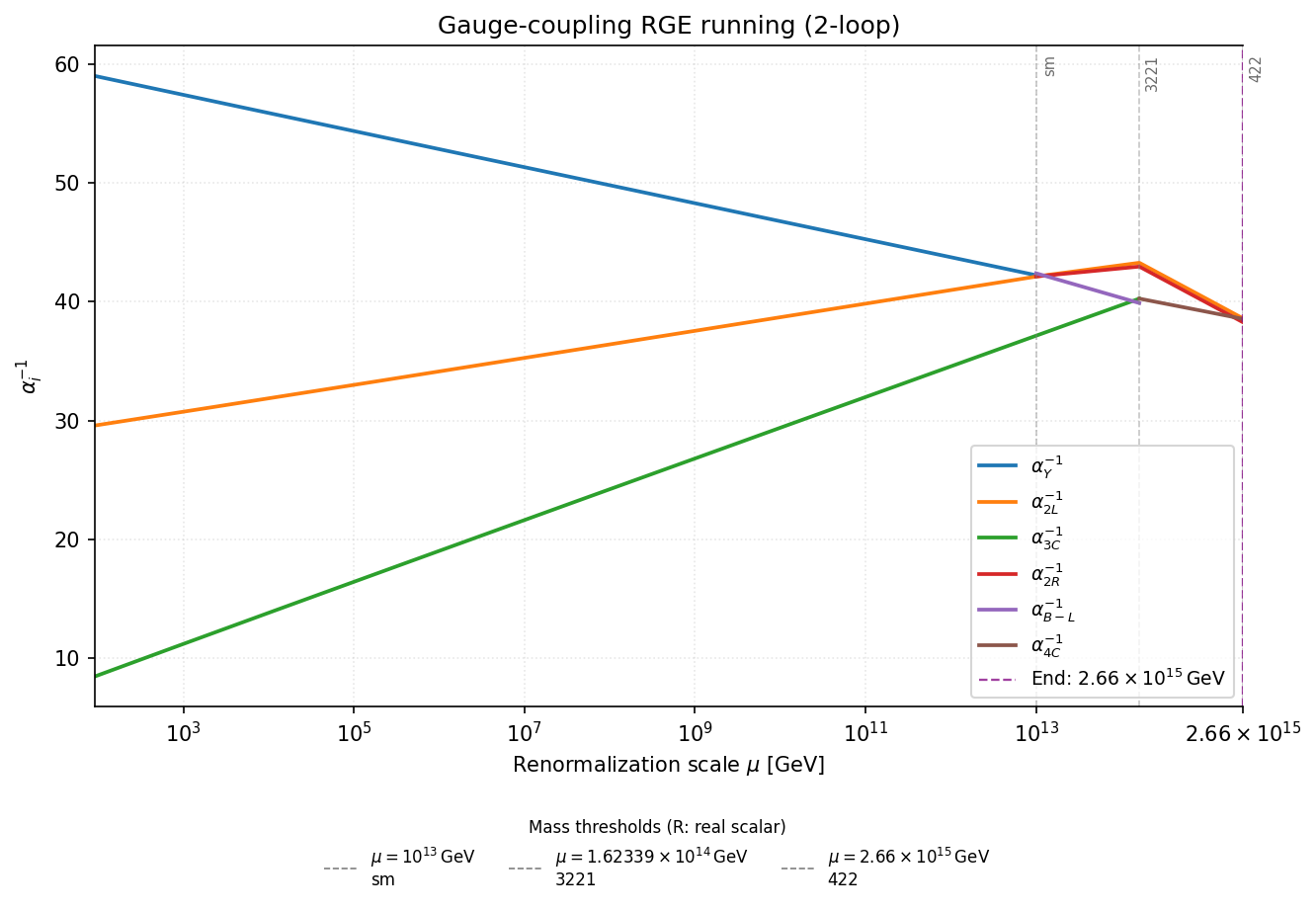}
\end{ProtonAgent}

%%%%%%%%%%%%%%%%%%%%%%%%%%%%%%%%%%%%%%%%%%%%%%%%%%%%%%
%%%%%%%%%%%%%%%%%%%%%%%%%%%%%%%%%%%%%%%%%%%%%%%%%%%%%%
\subsection{Proton-decay predictions and parameter scans}
\label{sec:examples:PD}

For proton decay calculations, the user may provide an input file containing all the parameters for a given benchmark point. The user then specifies the location of this file through natural-language instructions. As an example, the provided file \verb|input/nonsusy-SO10.txt| contains a benchmark parameter point consistent with the fermion mass fits of Ref.~\cite{Saad:2026ucf}. Here, the values of $M_\mathrm{GUT}$ and $g_\mathrm{GUT}$ are chosen independently for demonstration purposes.
\vspace{-8pt}\begin{User}
 an input file is given in \verb|input/nonsusy-SO10.txt|, can you compute the proton lifetime
\end{User}\vspace{-23pt}
\begin{ProtonAgent}
\centering
\includegraphics[width=0.65\textwidth]{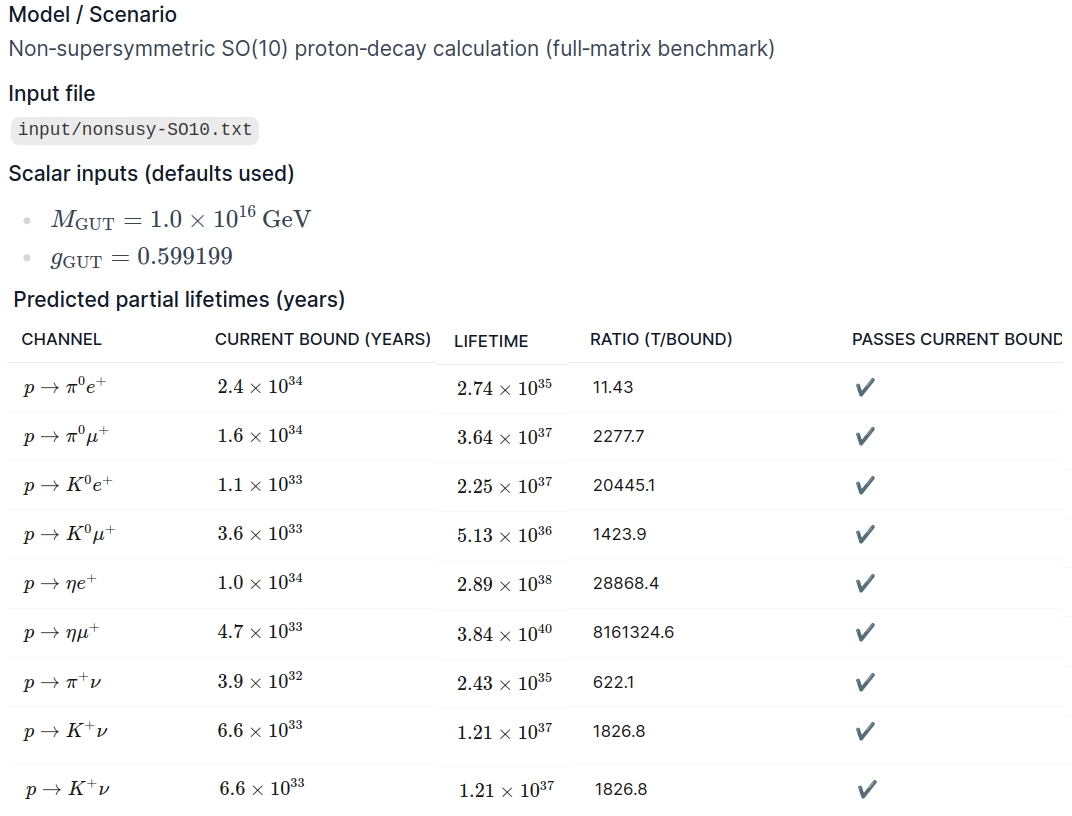}
\end{ProtonAgent}
\noindent As can be seen from this output of the agent, this specific parameter point satisfies all existing current bounds on proton decay lifetimes.

The agent provides a random-scan tool that generates random unitary matrices using the parametrizations described in Sec.~\ref{sec:pd}. For Example:
\vspace{-8pt}\begin{User}
 for this benchmark, run a non-SUSY SO(10) random scan of 500 points. Generate new random mixing matrices as well as random MGUT and gGUT. MGUT witin 1e15 to 1e16 GeV and gGUT in the range 0.4 to 0.8
\end{User}\vspace{-22pt}
\begin{ProtonAgent}
\centering
\includegraphics[width=0.65\textwidth]{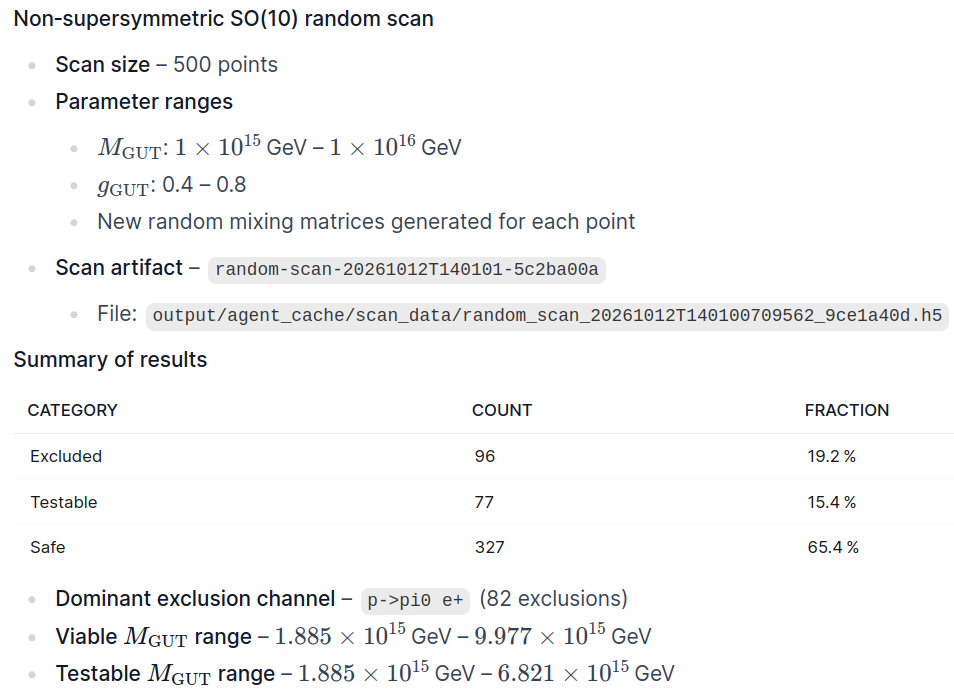}
\end{ProtonAgent}
Once the scan is completed, the user can request various plots to visualize and analyze the results. For example (In all illustrative examples presented here, ``X'' should be replaced with the corresponding actual file name. Explicitly specifying the correct file is important to prevent the agent from inadvertently selecting a different file, which could lead to incorrect results.):
\vspace{-8pt}\begin{User}
From scan file \verb|output/agent_cache/scan_data/random_scan_X.h5|, plot the partial lifetime for the most constraining channel versus MGUT. Use a logarithmic lifetime axis
\end{User}\vspace{-23pt}
\begin{ProtonAgent}
\centering
\includegraphics[width=0.65\textwidth]{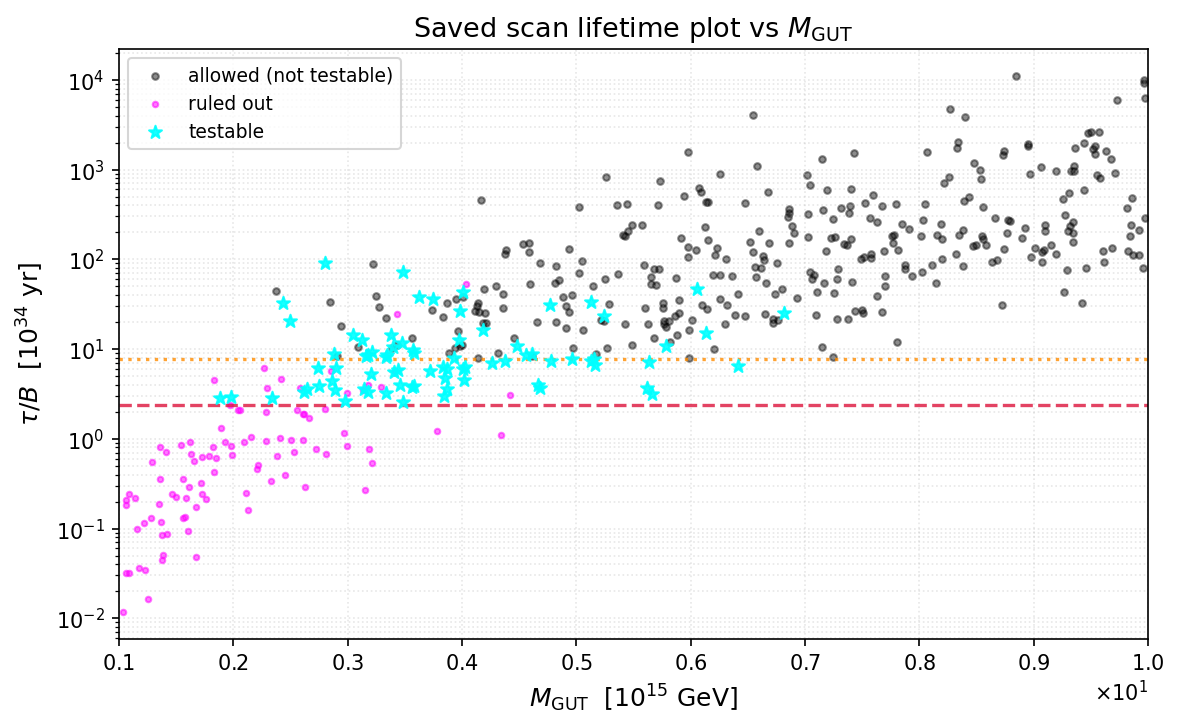}
\end{ProtonAgent}
In the above plot, the different colors represent the classification of parameter points based on their experimental constraint scores 
as mentioned in Sec.~\ref{sec:exp}.

\vspace{-8pt}\begin{User}
 From scan file \verb|output/agent_cache/scan_data/random_scan_X.h5|, plot the partial lifetime for all channels versus MGUT; both axes in log scales. 
\end{User}\vspace{-23pt}
\begin{ProtonAgent}
\centering
\includegraphics[width=1\textwidth]{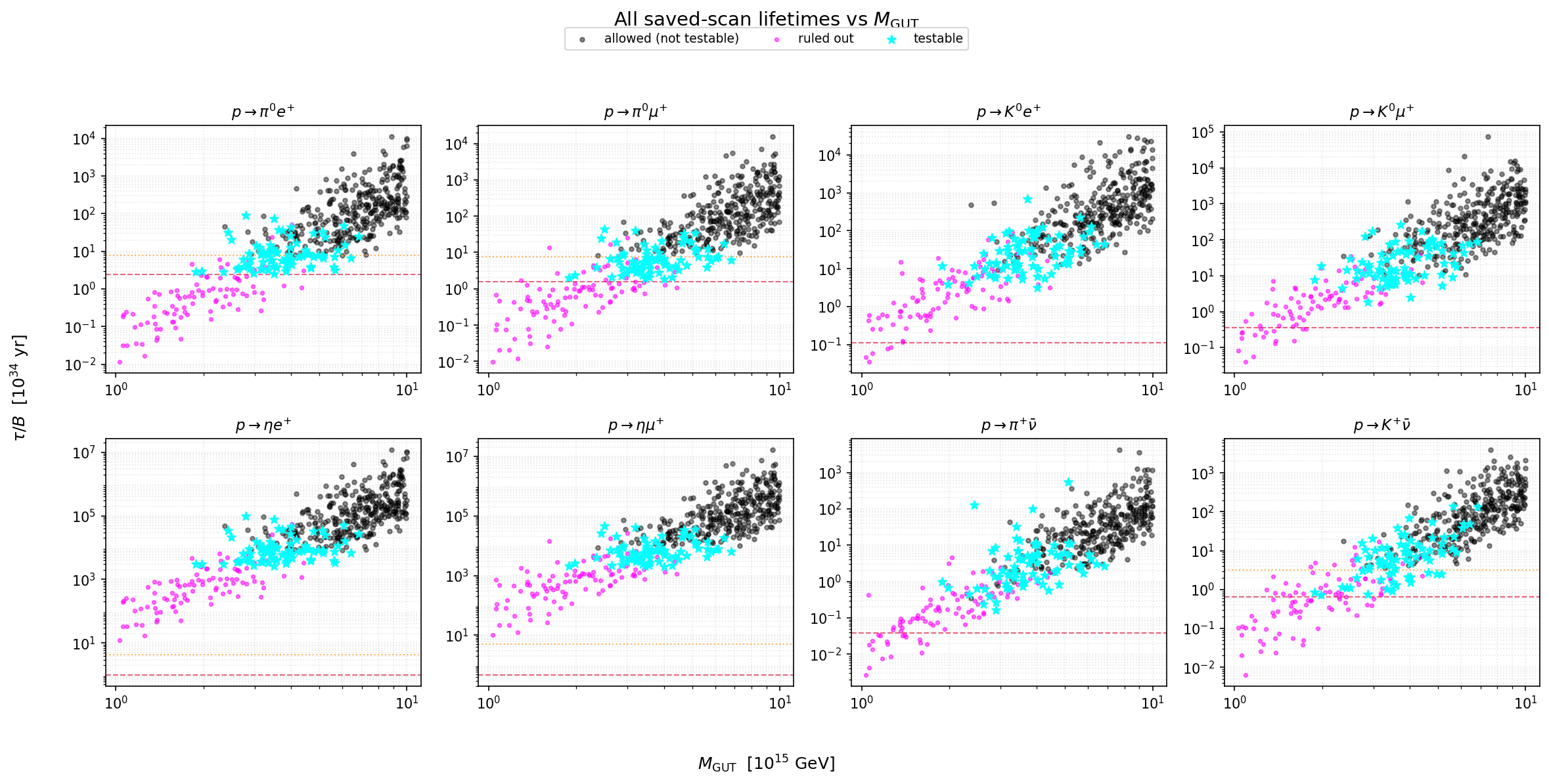}
\end{ProtonAgent}

To ensure scientific reproducibility, the agent automatically saves all relevant data generated during each computation. Users can also request a list of the quantities stored in the output files. In the following example, only a representative subset of the stored quantities is shown for brevity.
\vspace{-12pt}\begin{User}
 for the scan file \verb|output/agent_cache/scan_data/random_scan_X.h5|, what are the available columns?
\end{User}\vspace{-23pt}
\begin{ProtonAgent}
\centering
\includegraphics[width=0.4\textwidth]{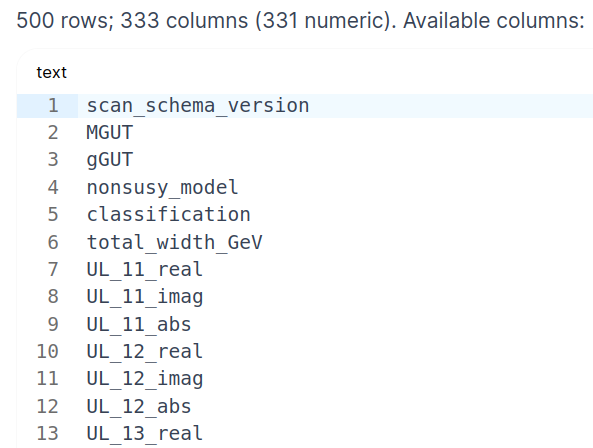}
\end{ProtonAgent}

Users can further request plots of the stored quantities to investigate parameter dependencies and their impact on proton decay predictions. For example:
\vspace{-8pt}\begin{User}
 From scan file \verb|output/agent_cache/scan_data/random_scan_X.h5|, plot the log scale of the lifetime of \verb|p→ pion^0 e^+| versus \verb|UL_12_abs|
\end{User}\vspace{-23pt}
\begin{ProtonAgent}
\centering
\includegraphics[width=0.65\textwidth]{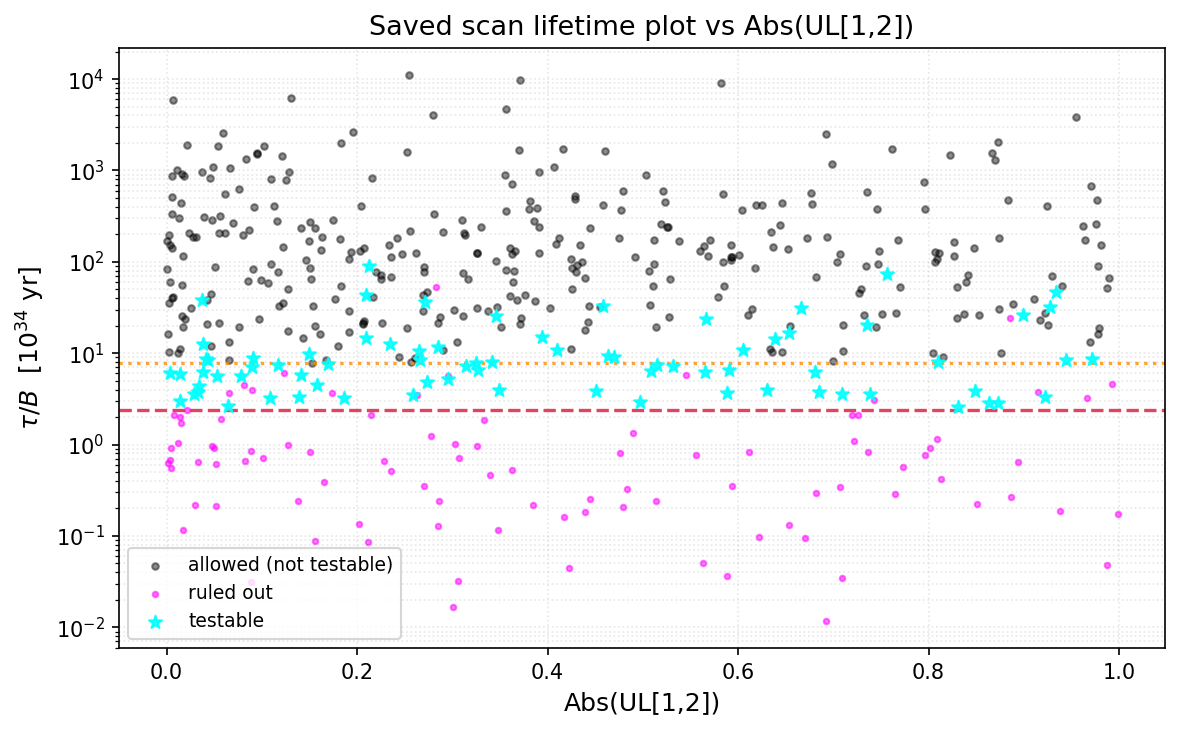}
\end{ProtonAgent}

Users can further request plots of the stored quantities to investigate parameter dependencies and their impact on proton decay predictions. They can also query the agent about the explicit dependence of proton decay rates or lifetimes on specific model parameters. For example:
\vspace{-8pt}\begin{User}
 from the plot you just made:  \verb|output/agent_cache/plots/lifetime_vs_UL12_abs.png|, explain the correlation of proton lifetime with  the parameter \verb|UL_12_abs|
\end{User}\vspace{-23pt}
\begin{ProtonAgent}
\centering
\includegraphics[width=0.7\textwidth]{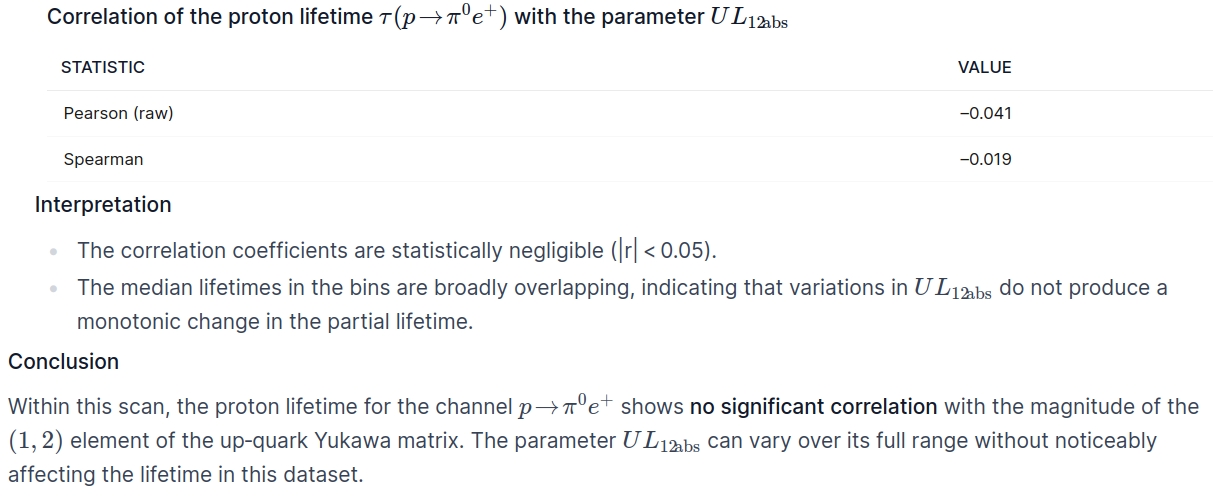}
\end{ProtonAgent}

Alternatively, users can investigate the dependence of proton decay predictions on other model parameters. For example:
\vspace{-8pt}\begin{User}
 from the dataset: \verb|output/agent_cache/scan_data/random_scan_X.h5|, explain the correlation of proton lifetime with the GUT scale
\end{User}\vspace{-23pt}
\begin{ProtonAgent}
\centering
\includegraphics[width=0.75\textwidth]{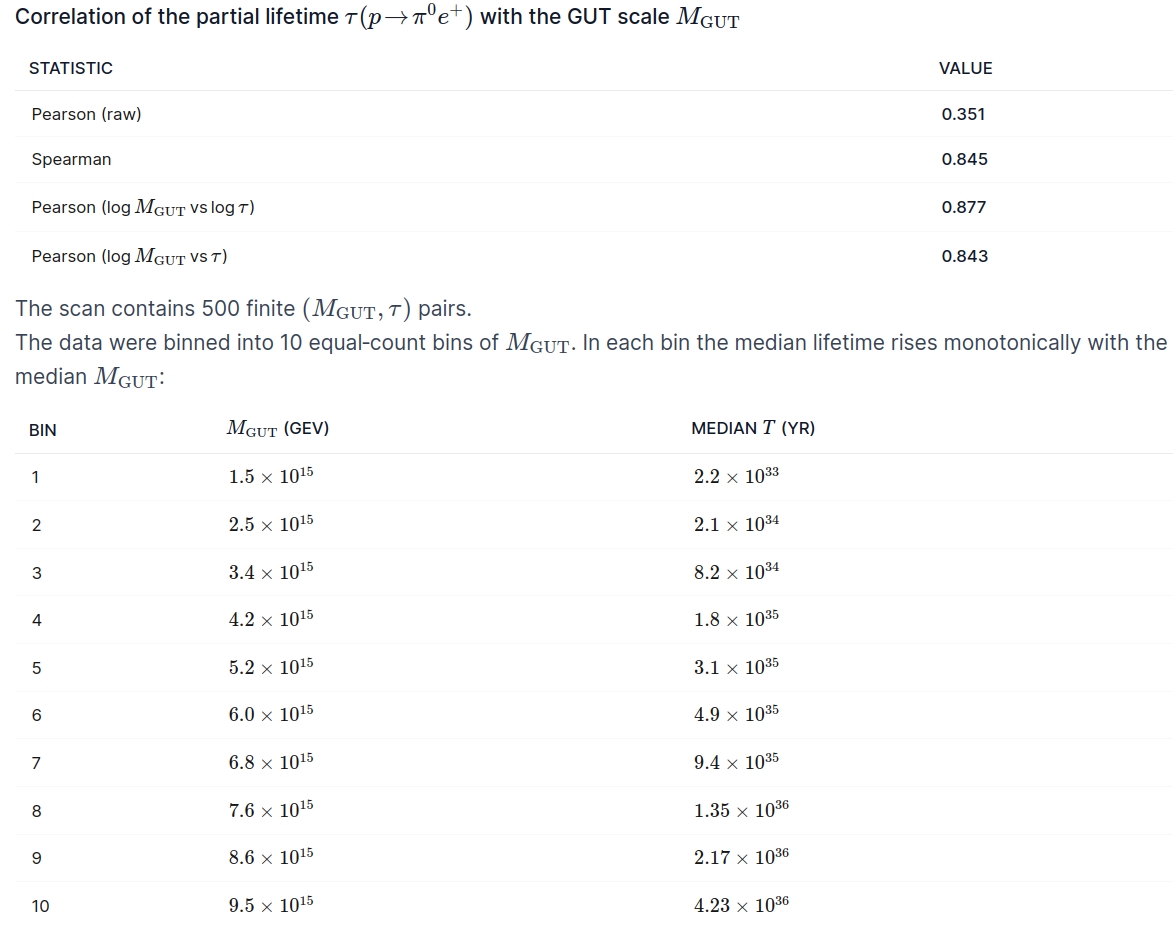}
\includegraphics[width=0.75\textwidth]{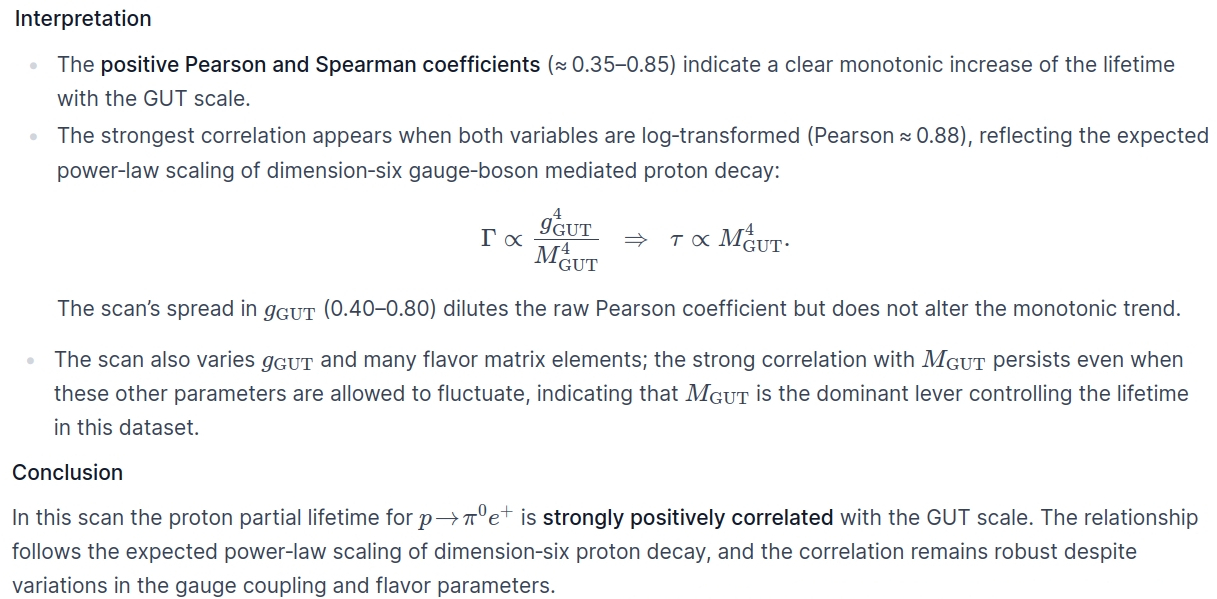}
\end{ProtonAgent}

Although stringent proton decay bounds might naively suggest that
the GUT scale must be close to $10^{16}$ GeV, lower scales can
remain viable if the relevant Wilson coefficients are sufficiently
suppressed~\cite{Nandi:1982ew}.
Here, we investigate how far the GUT scale can be lowered
in the Georgi--Glashow $SU(5)$ model (see, e.g.,
Ref.~\cite{Dorsner:2024seb}).
Our findings show that a GUT scale as low as
$5\times 10^{13}$ GeV can still be compatible with current
limits on gauge-mediated proton decay.
We use this benchmark to illustrate selected capabilities of
\PD; however, users are free to provide their own benchmark
parameter points for analysis.
\vspace{-8pt}\begin{User}
an input file is given in \verb|input/nonsusy-SU5.txt|, can you compute the proton lifetime
\end{User}\vspace{-23pt}
\begin{ProtonAgent}
\centering
\includegraphics[width=0.65\textwidth]{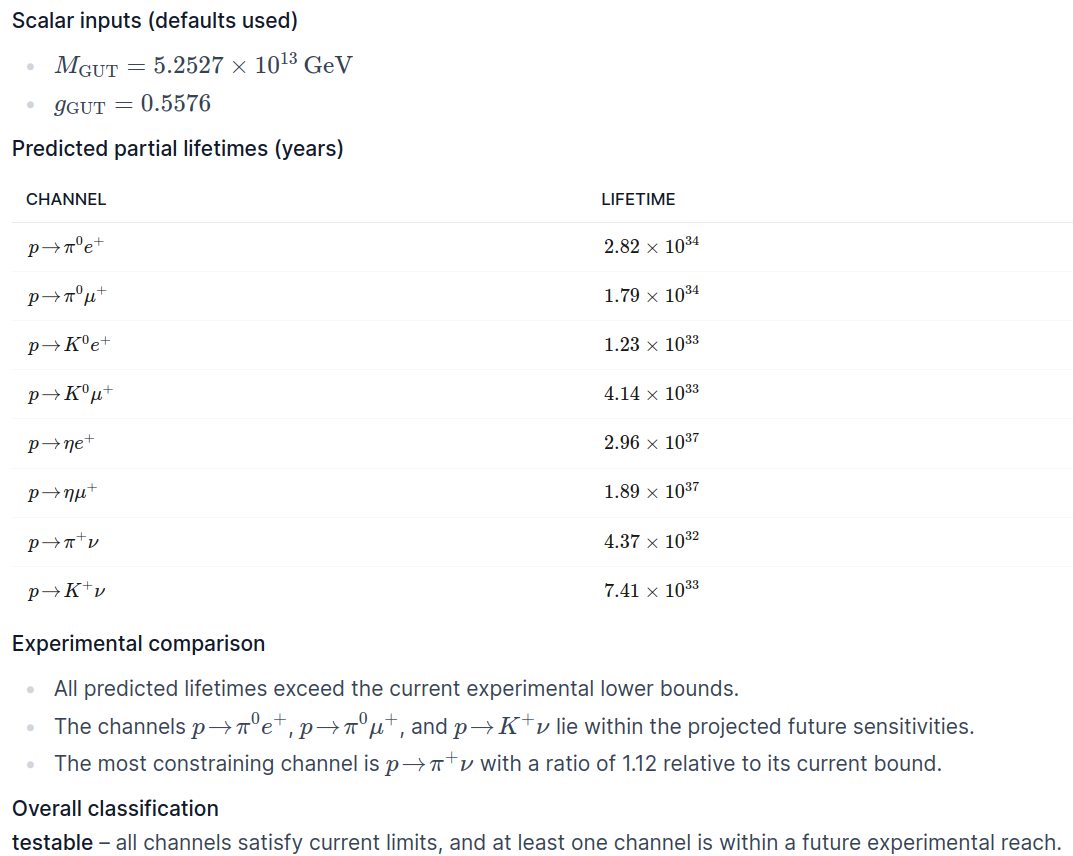}
\end{ProtonAgent}
\vspace{-8pt}\begin{User}
expand the description of the results obtained
\end{User}\vspace{-23pt}
\begin{ProtonAgent}
\centering
\includegraphics[width=0.65\textwidth]{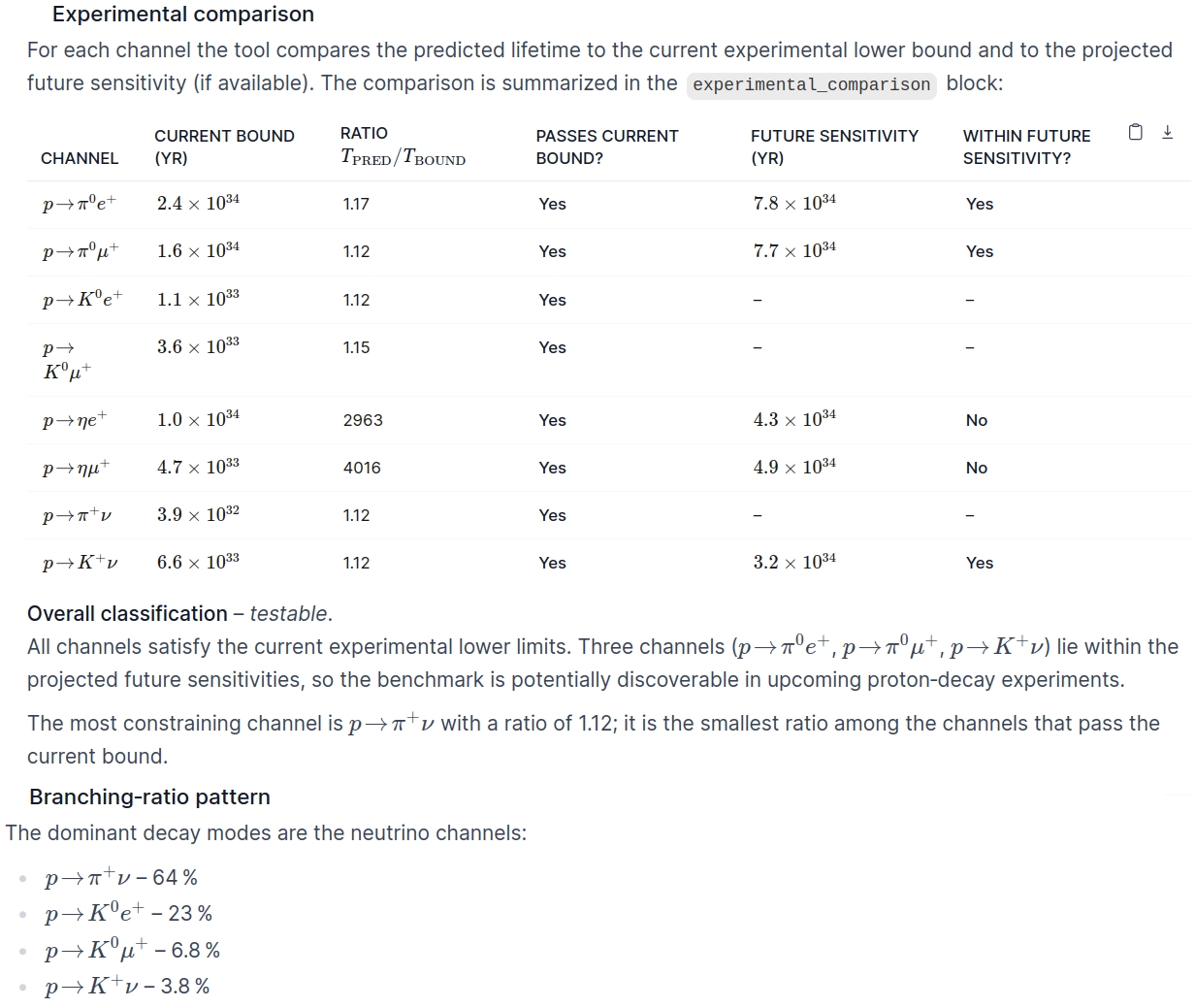}
\end{ProtonAgent}

Starting from a user-provided benchmark point, the agent can perform a Markov-chain Monte Carlo (MCMC) analysis and generate plots to explore the parameter space and its implications for proton decay predictions. Some examples are given in the following:
\vspace{-8pt}\begin{User}
starting from the input file \verb|input/nonsusy-SU5.txt|, perform a mcmc, 100000 points. vary the flavor mixing parameters by \verb|1e-5| but by keeping the MGUT and gGUT fixed
\end{User}\vspace{-23pt}
\begin{ProtonAgent}
\centering
\includegraphics[width=0.9\textwidth]{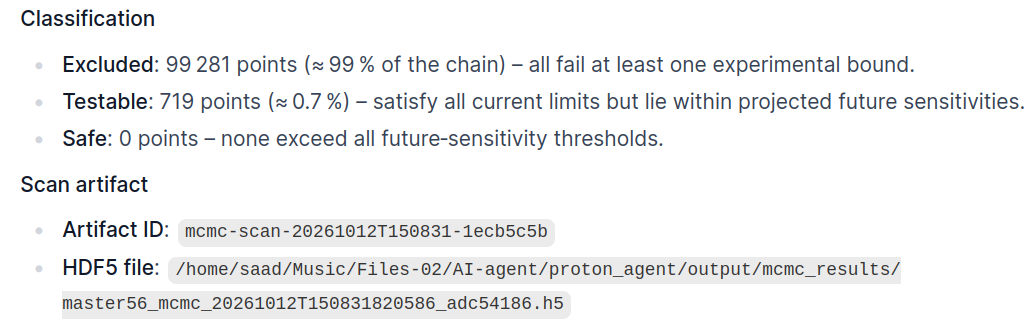}
\end{ProtonAgent}
\vspace{-8pt}\begin{User}
using this mcmc result: \verb|output/mcmc_results/master56_mcmc_X.h5 data|, plot lifetimes of all channels as a function of \verb|p→ pion^0 e^+|. therefore, it will be eight subplot.  both axes in log scale
\end{User}\vspace{-23pt}
\begin{ProtonAgent}
\centering
\includegraphics[width=0.65\textwidth]{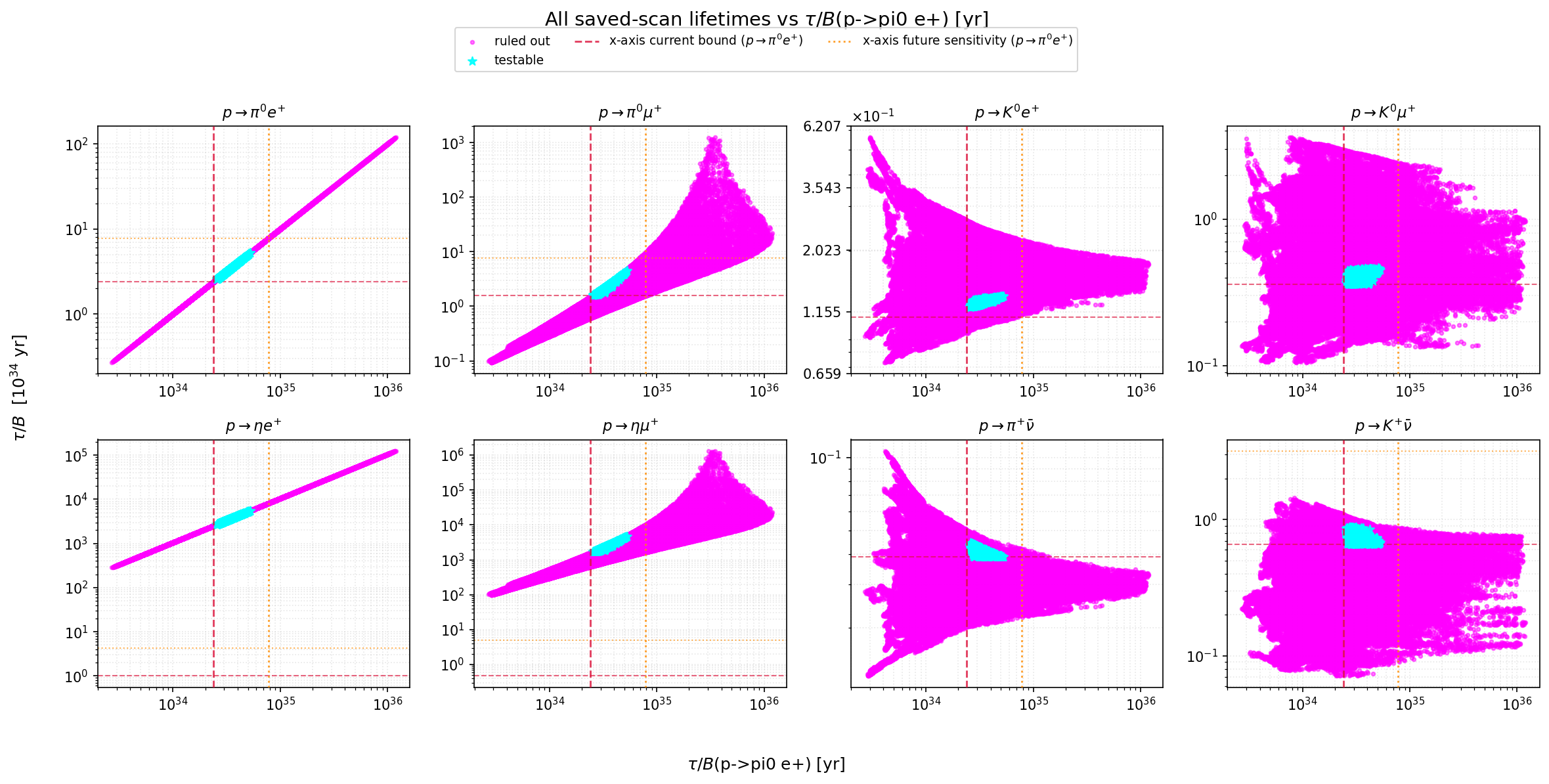}
\end{ProtonAgent}
\noindent Although such solutions require extreme fine-tuning, the MCMC analysis performed with 100,000 samples still identifies a small number of viable parameter points (shown in cyan) that satisfy the current experimental constraints.

In supersymmetric grand unified theories, the dominant contributions to proton decay are generally expected to arise from dimension-five operators mediated by color-triplet Higgs fields. In the following, we consider the provided benchmark point that is excluded by the experimental limits on all proton decay channels considered.
\vspace{-8pt}\begin{User}
 compute the proton lifetime for the file:  \verb|input/susy-SO10.txt|
\end{User}\vspace{-23pt}
\begin{ProtonAgent}
\centering
\includegraphics[width=0.65\textwidth]{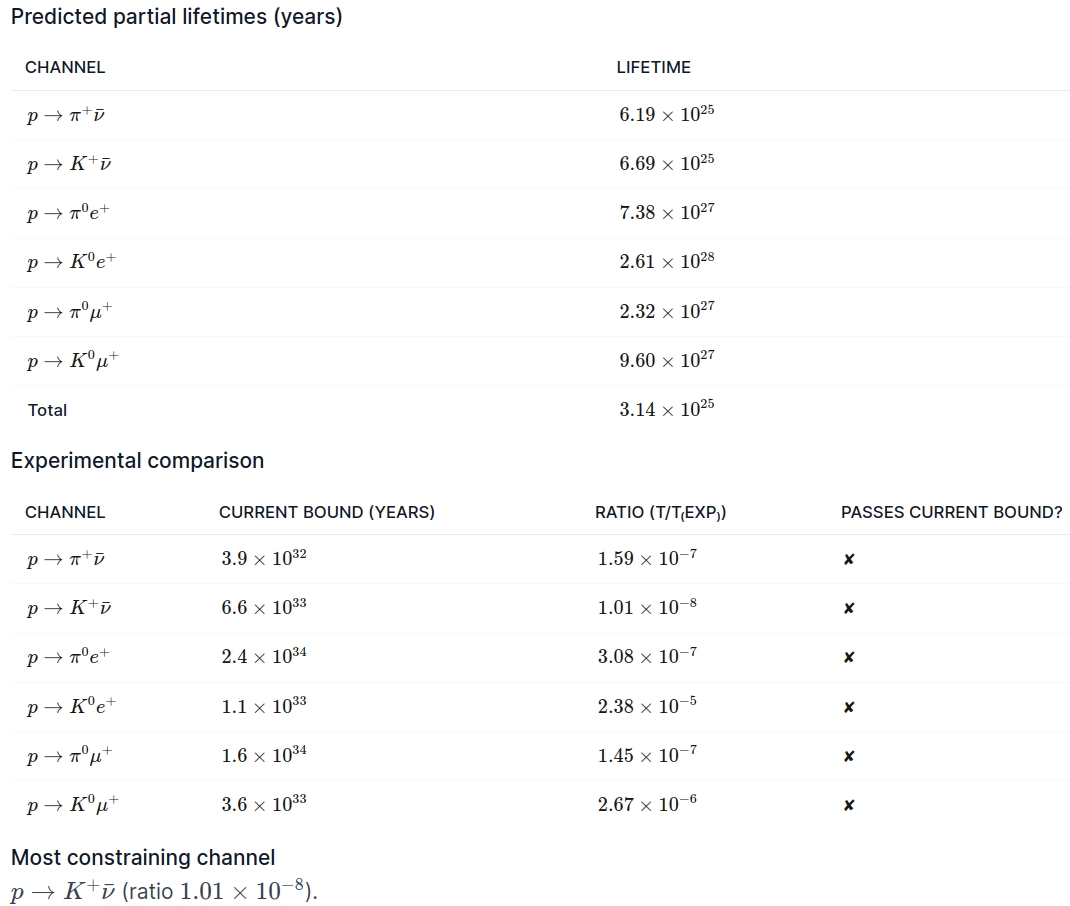}
\end{ProtonAgent}
As shown in the output, the most constraining proton decay channel has a predicted lifetime of approximately $6\times 10^{25}$ years, about eight orders of magnitude below the current experimental lower bound. Users can further instruct the agent to vary the color-triplet Higgs and sfermion masses to investigate their effects on the proton lifetime while keeping the flavor structure fixed, and visualize their effects on the proton lifetime.
\vspace{-8pt}\begin{User}
starting from the input file \verb|input/susy-SO10.txt|, perform a mcmc, 500 points. Vary the color triplet mass and the sfermion mass by a 0.05
\end{User}\vspace{-23pt}
\begin{ProtonAgent}
\centering
\includegraphics[width=0.65\textwidth]{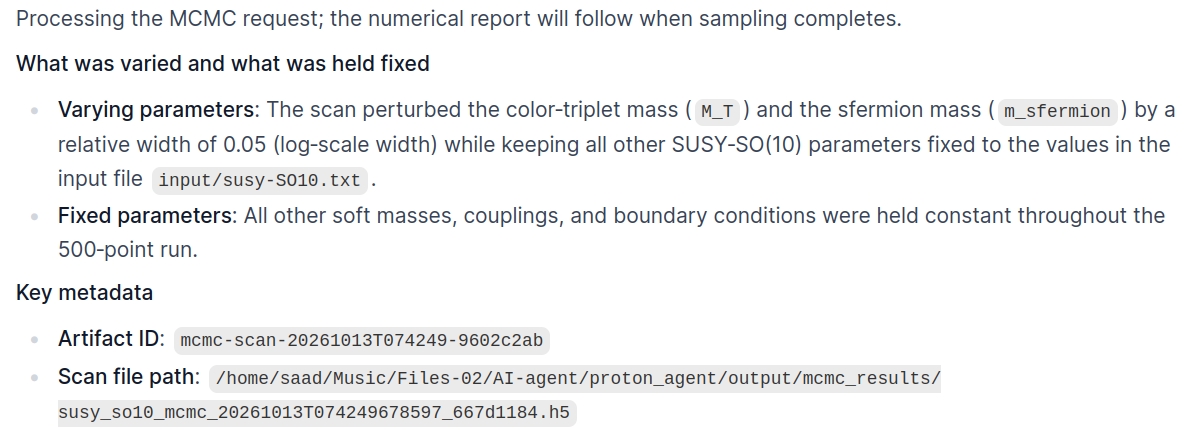}
\end{ProtonAgent}
\vspace{-8pt}\begin{User}
From \verb|output/mcmc_results/susy_so10_mcmc_X.h5| plot color triplet mass verses the sfermion mass, moreover, show heatmap of the lifetime of proton decay to kaon and anti-neutrino
\end{User}\vspace{-23pt}
\begin{ProtonAgent}
\centering
\includegraphics[width=0.65\textwidth]{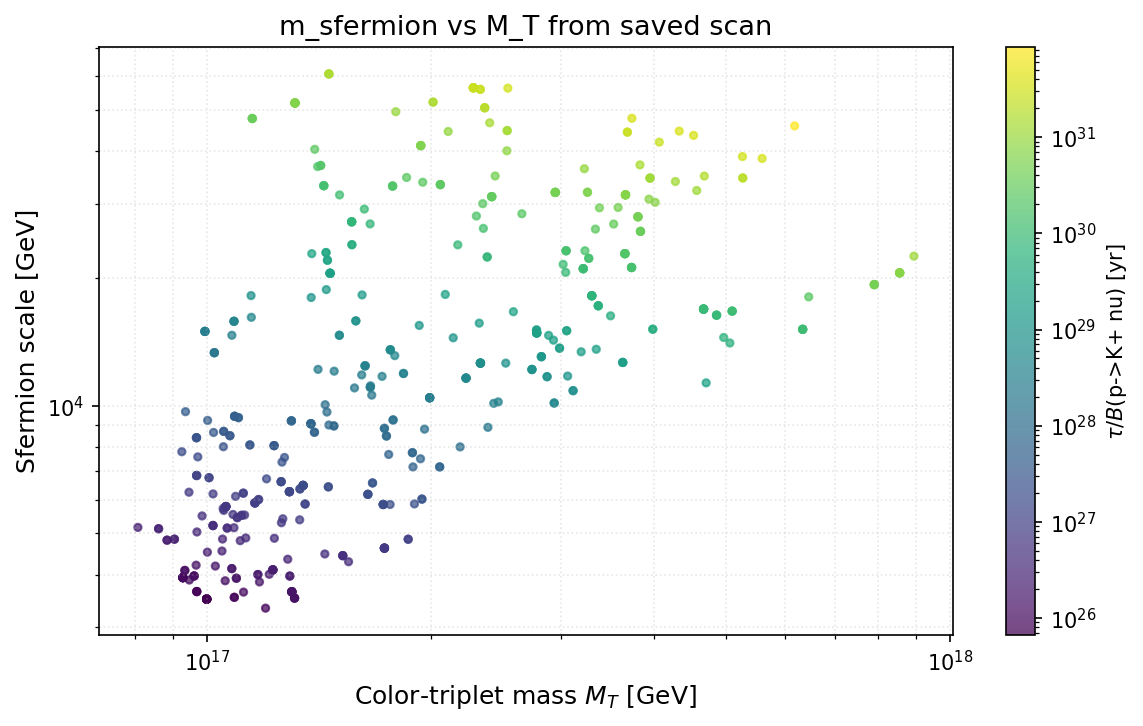}
\end{ProtonAgent}
\vspace{-8pt}\begin{User}
 without changing the flavor structure, meaning, the Yukawa and mixing matrices, how to make the given point \verb|input/susy-SO10.txt| compatible with the current experimental bounds
\end{User}\vspace{-23pt}
\begin{ProtonAgent}
\centering
\includegraphics[width=0.7\textwidth]{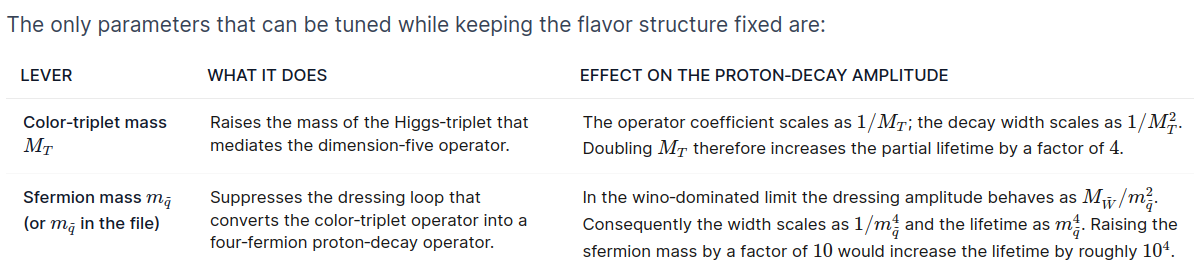}
\end{ProtonAgent}
Importantly, users can also query the agent about possible ways to evade the experimental proton decay bounds for a given benchmark point. For example:
\vspace{-8pt}\begin{User}
 in addition to keeping the flavor structure fixed as well as color-triplet mass fixed, for the given point \verb|input/susy-SO10.txt|,  determine the required sfermion mass scale to evade current experimental bounds
\end{User}\vspace{-23pt}
\begin{ProtonAgent}
\centering
\includegraphics[width=0.7\textwidth]{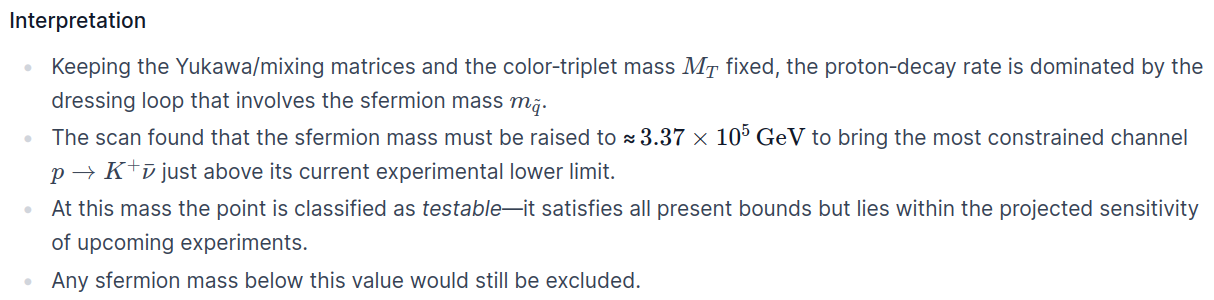}
\includegraphics[width=0.7\textwidth]{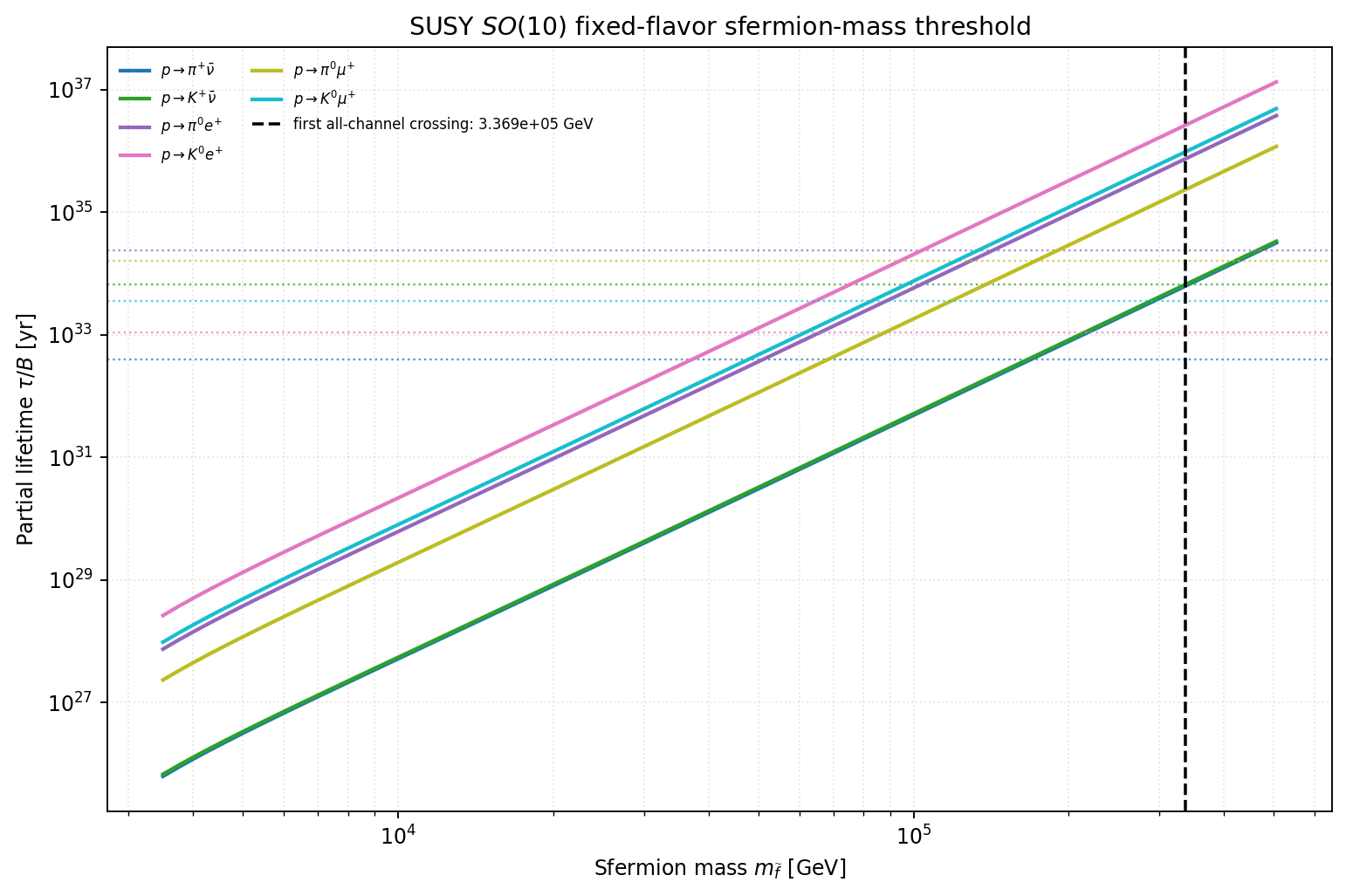}
\end{ProtonAgent}
\noindent In response to this specific query, the agent computes the lower bound on the sfermion mass required to satisfy the current experimental limits on proton decay.

In the case of SUSY \(SU(5)\), there are two phases \((\phi_{13}, \phi_{23})\) between the Yukawa couplings in the colored Higgs interactions, which cannot be removed by field redefinitions~\cite{Ellis:1979hy}. One can also investigate their dependence on the proton lifetime. For example:
\vspace{-8pt}\begin{User}
starting from the input point \verb|input/susy-SU5.txt|, make a scan of 100 points with \verb|phi13| and \verb|phi23| phases varied in the entire range 0 to \verb|2*pi|
\end{User}\vspace{-8pt}
After the run, a graphical view can be requested (as before, a correct file name must be used):
\vspace{-23pt}\begin{User}
make a plot of  lifetimes of \verb|p→ kaon + anti-neutrino| as a function of \verb|phi23| phase using the data: \verb|output/scan_results/susy_su5_phase_scan_X.h5|
\end{User}
\vspace{-23pt}
\begin{ProtonAgent}
\centering
\includegraphics[width=0.65\textwidth]{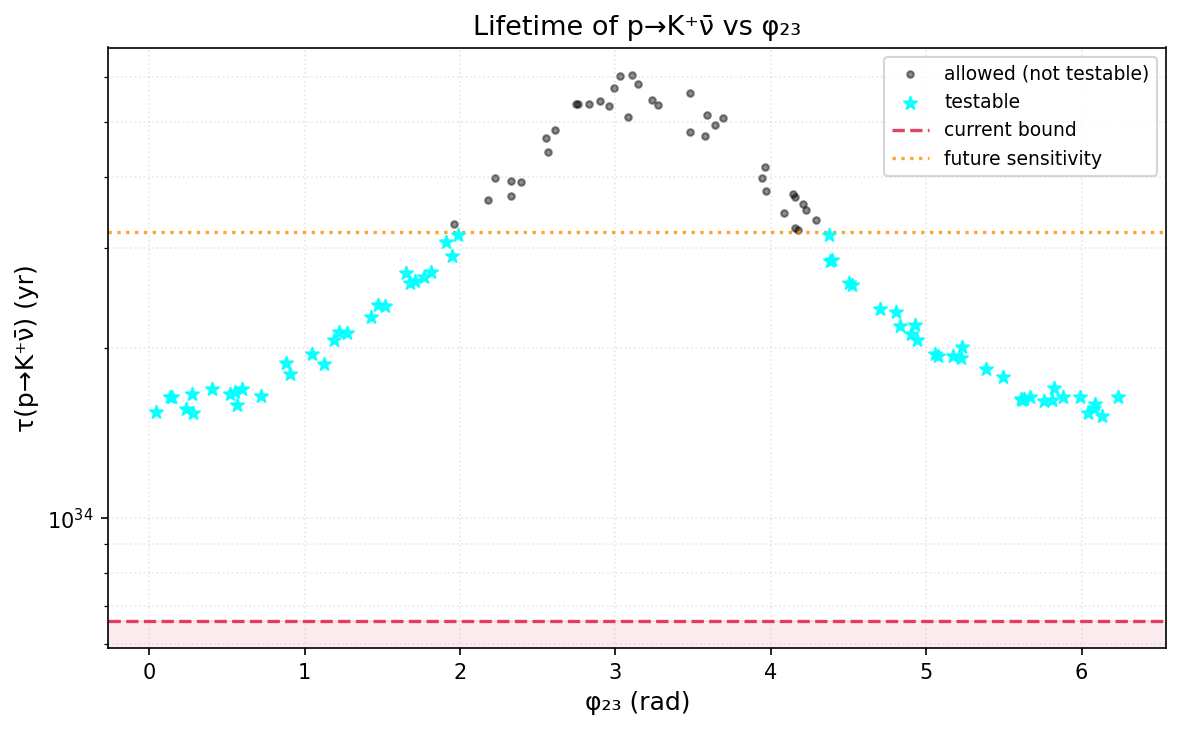}
\end{ProtonAgent}

The illustrative examples presented above demonstrate the usefulness of this domain-specific AI agent in assisting researchers with accurate scientific computations, systematic parameter-space scans, MCMC analyses, and the extraction of physical insights into various aspects of gauge coupling unification and proton decay.

%%%%%%%%%%%%%%%%%%%%%%%%%%%%%%%%%%%%%%%%%%%%%%%%%%%%%%
\section{Conclusions}
\label{sec:conclusion}

We have presented \PD, a tool-grounded AI agent for natural-language
interaction with gauge coupling unification and proton decay calculations.
The framework computes one- and two-loop gauge beta coefficients for
supported representations and evolves couplings through supported
symmetry-breaking stages. It evaluates gauge-mediated dimension-six
proton decay in non-supersymmetric $SU(5)$ and $SO(10)$ and dressed
dimension-five contributions in their supersymmetric counterparts.
Benchmark calculations, parameter scans, MCMC exploration, comparisons
with experimental limits, and visualizations provide complementary
ways to study these models, enabling researchers to investigate
parameter dependencies, interpret numerical results, and gain
physical insights into gauge coupling unification and proton decay.

The language model interprets user requests, selects computational
tools, and assists in the physical interpretation of their outputs,
while dedicated scientific routines perform the underlying
calculations. LangGraph coordinates these workflows, and saved
results support inspection, reproducibility, and reuse.
Nevertheless, independent physics validation and human judgment
remain essential. The code is publicly available at  \href{https://github.com/saad-hep/proton_agent}{\faGithub~proton\_agent}.

Scientific AI agents are still at an early stage of development,
but their capabilities are expected to grow rapidly with advances
in language models and scientific computational tools.
Beyond extending the physics coverage of individual agents,
an interesting future direction is to integrate multiple
domain-specific scientific AI agents into a unified multi-agent
framework, where specialized agents collaborate on different
aspects of theoretical model building and phenomenological
analysis. Such an approach could enable more comprehensive and
automated investigations of beyond the Standard Model physics.

%%%%%%%%%%%%%%%%%%%%%%%%%%%%%
\section*{Acknowledgments}
%%%%%%%%%%%%%%%%%%%%%%%%%%%%%
OpenAI's GPT-5.6-Sol and GPT-6.1-Sol models were used to assist
with code generation and the preparation of parts of the manuscript.
However, the underlying scientific computational tools  were developed by the author independently of these language models. S.S.\ acknowledges financial support from the Slovenian Research Agency (research core funding No.\ P1-0035).

\appendix
\numberwithin{equation}{section}
%%%%%%%%%%%%%%%%%%%%%%%%%%%%%%%%%%%%%%%%%%%%%%%%%%%%%%
%%%%%%%%%%%%%%%%%%%%%%%%%%%%%%%%%%%%%%%%%%%%%%%%%%%%%%
\section{Input conventions and run options}
\label{app:interface_reference}

This appendix collects the default inputs, conventions, and run options
for gauge-coupling evolution and proton-decay calculations through
Open WebUI. The tables distinguish direct chat settings from
server-accessible input-file values and specify the parameter variations
available in scans and MCMC.

\subsection{Gauge-coupling evolution and unification}
\label{sec:rge_input_values}

\paragraph{Boundary values and loop order.}
Gauge-coupling evolution uses one loop by default. The default initial
scale and gauge couplings are~\cite{Antusch:2013jca}
\begin{equation}
\mu_0=M_Z=91.1876~\mathrm{GeV},\qquad
(g_1,g_2,g_3)(\mu_0)=(0.461425,\,0.65184,\,1.2143),
\label{eq:rge_boundary_inputs}
\end{equation}
where $g_1=\sqrt{5/3}\,g_Y$. These input values can be changed by the user, see Table~\ref{tab:rge_agent_inputs}. For example, the user may prefer to use more up-to-date values as given in Ref.~\cite{Antusch:2025fpm}.  For two-loop single-group SM running,
the same gauge boundary values are used together with the Yukawa
inputs~\cite{Antusch:2013jca}
\begin{align}
(y_t,y_b,y_\tau)(\mu_0)&=(0.9861,\,0.01639,\,0.010029).
\label{eq:rge_yukawa_inputs}
\end{align}
These input values can be changed by the user, see Table~\ref{tab:rge_agent_inputs}. 
Only these third-generation Yukawa couplings are evolved, using
one-loop Yukawa equations; their contribution to gauge evolution
enters at two loops. First- and second-generation Yukawa couplings and
additional BSM Yukawa interactions are omitted.
The staged non-SUSY and SM-to-MSSM workflows
use gauge-only running at both orders and do not use
Eq.~\eqref{eq:rge_yukawa_inputs}.

\paragraph{Running intervals and matching.}
For single-group SM and SM-to-MSSM running, the default final scale
is $10^{16}~\mathrm{GeV}$. This is a user-adjustable integration
endpoint; a common intersection of the couplings is not imposed.
The SM-to-MSSM workflow uses the matching scale
$M_{\rm SUSY}=10^3~\mathrm{GeV}$ by default and converts the
$\overline{\rm MS}$ gauge couplings to $\overline{\rm DR}$ there.
It requires $\mu_0<M_{\rm SUSY}<\mu_{\rm end}$.
For a staged non-SUSY run, all stage endpoints are supplied by the
user in increasing order, and the last endpoint fixes the final scale.
Matching between different non-SUSY gauge groups is tree-level;
finite threshold corrections and Abelian kinetic mixing are omitted.

\paragraph{Particle content and thresholds.}
The SM and MSSM tools include their respective baseline particle
contents. Intermediate-group tools include their predefined matter
contents by default. Additional multiplets are model
inputs, with multiplicity one by default. Non-SUSY scalars default
to complex and fermions to Weyl; the MSSM field parser defaults to a
chiral superfield. A supplied mass activates an extra multiplet at its
threshold, while an omitted mass makes it active throughout the interval.
No common default mass is assigned to extra multiplets.

\paragraph{Calculation and plotting options.}
Table~\ref{tab:rge_agent_inputs} summarizes requests for beta coefficients,
coupling evolution, unification diagnostics, and plots. Multiplet notation
and representation limits are given in Table~\ref{tab:beta_representations}.

\begingroup
\small
\renewcommand{\arraystretch}{1.15}
\setlength{\tabcolsep}{5pt}
\begin{longtable}{
>{\raggedright\arraybackslash}p{0.24\textwidth}
>{\raggedright\arraybackslash}p{\dimexpr0.76\textwidth-4\tabcolsep\relax}}
\caption{Gauge-coupling calculation options available.
The entries describe chat requests rather than internal tool argument
schemas. Matching prescriptions, charge normalizations, and solver
tolerances are fixed implementation choices.}
\label{tab:rge_agent_inputs}
\\
\toprule
\textbf{Chat request} & \textbf{What the user specifies and its scope}\\
\midrule
\endfirsthead
\multicolumn{2}{l}{\textit{Table~\thetable\ continued}}\\
\toprule
\textbf{Chat request} & \textbf{What the user specifies and its scope}\\
\midrule
\endhead
\midrule
\multicolumn{2}{r}{\textit{Continued on next page}}\\
\endfoot
\bottomrule
\endlastfoot
Calculation type &
Ask for SM, MSSM, or intermediate-group beta coefficients; SM or
SM-to-MSSM running; a supported non-SUSY symmetry chain; or a running plot.
\\
\midrule
Loop order &
Request one-loop or two-loop running explicitly; the default is one loop.
The Yukawa treatment follows Sec.~\ref{sec:rge_input_values}.\\
\midrule
Running interval &
State the starting and final scales in GeV, for example,
``from $M_Z$ to $10^{15}$ GeV.'' Omitted values use the defaults above;
for a symmetry chain, state each transition scale and the final endpoint.\\
\midrule
Initial gauge couplings &
Include explicit assignments such as \code{g1=0.46}, \code{g2=0.65},
\code{g3=1.2}, together with their reference scale.
Otherwise Eq.~\eqref{eq:rge_boundary_inputs} supplies the defaults.
For another starting scale, supply the corresponding boundary values. For two-loop scenario, the starting values of the SM third-generation Yukawa couplings can also be specified such as \code{yt=0.8}, \code{yb=0.01}, \code{ytau=0.095}; default values of Eq.~\eqref{eq:rge_yukawa_inputs} are used otherwise.\\
\midrule
Additional SM multiplets &
State the scalar or fermion type, the representation tuple from
Table~\ref{tab:beta_representations}, and each activation mass.
Complex scalars and Weyl fermions are the defaults.\\
\midrule
Non-SUSY symmetry chain &
Specify the sequence of SM, LR/$G_{3221}$, $G_{3211}$,
PS/$G_{422}$, or $G_{421}$ stages, their transition scales,
and the extra multiplets present in each stage.
Group matching and the predefined baseline matter are applied automatically.\\
\midrule
SM-to-MSSM transition &
State the SUSY matching scale and final running scale, and any extra
SM multiplets below or chiral superfields above the matching scale.
The default matching scale is $10^3$ GeV.\\
\midrule
Output sampling &
Request a sample count, for example, ``400 points'' or
``400 points per stage.'' Defaults are 250 for an SM run and each
SM-to-MSSM segment, and 160 per staged non-SUSY interval.
The common sample-count arguments are capped at 2000.\\
\midrule
Plot quantity &
Ask for gauge couplings $g_i$ or explicitly request inverse couplings
$\alpha_i^{-1}=4\pi/g_i^2$. New running plots default to $g_i$;
a request to redisplay an existing plot retains its quantity.
\\
\midrule
SM unification diagnostic &
Ask whether specified SM multiplets help gauge unification.
This diagnostic is one-loop and distinguishes improvement from an
actual crossing. For one multiplet, a physical threshold mass and crossing
scale are returned when a solution exists. An explicit two-loop request
uses numerical evolution instead.\\
\end{longtable}
\endgroup

\subsection{Proton-decay inputs and analysis}
\label{sec:proton_input_controls}

\paragraph{Input conventions.}
The physical inputs are defined in Table~\ref{tab:proton_inputs}.
How those inputs are supplied
and varied are discussed below, while Table~\ref{tab:proton_chat_options} summarizes the
supported workflows. Example input files are provided in the repository's
\code{input/} directory:
\code{nonsusy-SU5.txt}, \code{nonsusy-SO10.txt},
\code{susy-SU5.txt}, and \code{susy-SO10.txt}.
Users may select a supplied file or provide their own by specifying
its path in the request. Paths refer to the agent server's filesystem.

Masses and scales are in GeV, lifetimes in years, and angles and phases
in radians. Couplings, Yukawa entries, mixing coefficients, and $A_X$
are dimensionless. Hadronic matrix elements and the stored
experimental-bound prescriptions are internal inputs rather than
independent chat controls.

\paragraph{Benchmark flavor values.}
\label{app:proton_benchmarks}
For the reduced non-SUSY $SU(5)$ input option in
Table~\ref{tab:proton_inputs}, the CKM and PMNS matrices are constructed
using the parameterization of Eq.~\eqref{eq:general_unitary}.
Their fixed benchmark inputs are
\begin{align}
(\theta_{12},\theta_{23},\theta_{13},\delta)_{\rm CKM}
&=(0.2270,\,0.04194,\,0.0037,\,1.139) \text{~\cite{Antusch:2025fpm}},\notag\\
(\theta_{12},\theta_{23},\theta_{13})_{\rm PMNS}
&=(0.589202,\,0.75538,\,0.150535) \text{~\cite{NUFIT,Esteban:2020cvm}},
\label{eq:reduced_flavor_mixing_inputs}
\end{align}
with all entries in radians. The five external CKM phases, the PMNS
Dirac phase, and the five external PMNS phases are supplied by the
user in the reduced input file and form part of its 56 flavor parameters.
Thus the full CKM and PMNS matrices are not fixed numerical matrices.
These fixed angle choices apply to the reduced file option; full-matrix
files supply $D$ and $N$ directly, while the separate constrained
random/grid scan mode permits its own CKM/PMNS settings.

For SUSY $SU(5)$~\cite{Goto:1998qg}, omitted Yukawa inputs use the following benchmark values~\cite{Antusch:2025fpm}, quoted at 1 TeV:
\begin{align}
&f_U=\operatorname{diag}\!\left(
6.09\times10^{-6},\;
3.04\times10^{-3},\;
0.8472
\right), \label{eq:fU}
\\
&f_D^\mathrm{diag}=
\operatorname{diag}\!\left(
1.33\times10^{-5},\;
2.66\times10^{-4},\;
1.373\times10^{-2}
\right),
\\
&f_E=\operatorname{diag}\!\left(
2.828\times10^{-6},\;
5.9576\times10^{-4},\;
1.0119\times10^{-2}
\right).     \label{eq:fE}
\end{align}

\paragraph{Input notation and file requirements.}
The inputs in Table~\ref{tab:proton_inputs} may be supplied through
supported chat arguments or benchmark files. Full non-SUSY flavor
matrices and complex SUSY $SU(5)$ Yukawa matrices require files.
Fixed-flavor scans retain the supplied flavor structure.
Non-SUSY MCMC supports variations of flavor and gauge inputs.
SUSY $SU(5)$ MCMC varies selected scalar parameters and phases,
with Yukawa matrices fixed; SUSY $SO(10)$ MCMC additionally
supports variations of Yukawa and mixing inputs.
The dressing gauge couplings and, for SUSY $SO(10)$,
the renormalization factors $A_L$ and $A_S$ can be specified
in benchmark files but are held fixed during scans and MCMC. For both the SUSY and non-SUSY frameworks, the running of the renormalization group equations is currently not explicitly implemented and has been left for future inclusion in the tools.

\paragraph{Calculation and analysis options.}
Table~\ref{tab:proton_chat_options} collects point calculations, scans,
MCMC settings, saved-data analysis, plotting, and experimental comparisons.

\begingroup
\small
\renewcommand{\arraystretch}{1.15}
\setlength{\tabcolsep}{5pt}
\begin{longtable}{
>{\raggedright\arraybackslash}p{0.24\textwidth}
>{\raggedright\arraybackslash}p{\dimexpr0.76\textwidth-4\tabcolsep\relax}}
\caption{Proton-decay workflows available through Open WebUI.
Physical inputs and their access routes are specified in
the text.}
\label{tab:proton_chat_options}\\
\toprule
\textbf{Workflow} & \textbf{Request and scope}\\
\midrule
\endfirsthead
\multicolumn{2}{l}{\textit{Table~\thetable\ continued}}\\
\toprule
\textbf{Workflow} & \textbf{Request and scope}\\
\midrule
\endhead
\midrule
\multicolumn{2}{r}{\textit{Continued on next page}}\\
\endfoot
\bottomrule
\endlastfoot
Point calculations &
Select a framework and parameter point or input file; request channel
lifetimes, widths, or branching ratios. Supported SUSY file-point
overrides retain all other inputs.\\
\midrule
Parameter scans &
Choose generated-flavor random or grid scans for non-SUSY models,
or a fixed-flavor file scan for any framework. Specify parameter ranges
and a sample count or grid size.\\
\midrule
MCMC exploration &
Specify an initial file or the last computed point, retained sample
count, varied parameters, and proposal widths. The code uses acceptance based on \(\exp[-\Delta R_{\max}/2]\) as a constraint-guided exploration score. \\
\midrule
Hadronic matrix elements& 
The user can update the hadronic matrix elements in the input file \verb|input/matrix-element.txt| and requests the agent to use it during computations. For example, ``Compute proton lifetimes using \verb|input/susy-SU5.txt|, and use \verb|input/matrix-elements.txt| for the hadronic matrix elements''.  Otherwise, default values are used.  
\\
\midrule
Experimental probes& 
The user can update the experimental constraints (current and future sensitivities) in the input file \verb|input/bounds.txt| and requests the agent to use it during computations. For example ``Compute proton lifetimes using \verb|input/susy-SU5.txt| and use \verb|input/bounds.txt| for experimental bounds and future sensitivities''. Otherwise, default values are used.  
\\
\midrule
Saved-data inspection &
Identify an HDF5 path or artifact ID; request available columns or
summary statistics. Inspection reuses the saved samples.\\
\midrule
Plots and correlations &
Request 2D or 3D plots of saved numeric columns, or single- and
all-channel lifetime plots. Axis scales and numeric coloring are
selectable. Relationship analysis returns Pearson and Spearman
correlations and binned summaries.\\
\midrule
Experimental comparisons &
Request stored experimental bounds, future sensitivities,
point classification, or comparisons of the factors controlling
different non-SUSY decay channels. For SUSY, a
minimum-sfermion-mass search holds the other file inputs fixed.\\
\end{longtable}
\endgroup

For file-based SUSY MCMC, \code{log\_width} is a Gaussian proposal width
in $\log_{10}$; a value of 0.05 does not denote a bounded $\pm5\%$
interval. Default MCMC settings are 500 retained samples, seed 24680,
zero burn-in, thinning one, and all available CPU cores, with separate
chains per worker. Lifetime coloring uses scatter plots, without heatmap
interpolation; logarithmic axes require positive values.

%%%%%%%%%%%%%%%%%%%%%%%%%%%%%%%%%%%%%%%%%%%%%
%%%%%%%%%%%%%%%%%%%%%%%%%%%%%%%%%%%%%%%%%%%%%
\bibliographystyle{style}
\bibliography{reference}
\end{document}